\documentclass[10pt,journal,compsoc,table]{IEEEtran}

\usepackage{epsfig}
\usepackage{epstopdf}
\usepackage{times}
\usepackage{framed}
\usepackage{enumitem}
\usepackage{float}
\usepackage{etoolbox}
\usepackage{algorithmic}
\usepackage{amssymb}
\usepackage{amsmath}
\usepackage{multirow}
\usepackage{url}
\usepackage{lipsum}
\usepackage{soul}
\usepackage{caption}
\usepackage{mathtools}
\usepackage{bbm}
\usepackage{booktabs}
\usepackage{xcolor}
\usepackage{makecell}
\usepackage{pifont}
\usepackage{subfigure}
\usepackage[export]{adjustbox} 
\usepackage[norelsize,boxruled]{algorithm2e}
\makeatletter
\newcommand{\removelatexerror} {\let\@latex@error\@gobble}
\makeatother

\newcommand{\change}[1]{{\color{black}{#1}}}

\def\1{\mathbf{1}}
\newcommand{\vv}[1]{\boldsymbol{#1}}
\newcommand{\vvi} [2]{ \boldsymbol{#1}^{(#2)} }

\newcounter{subeqn} %
\makeatletter
\@addtoreset{subeqn}{equation}
\makeatother

\usepackage{array} 
\def\ignore#1\endignore{}
\newcolumntype{h}{@{}>{\ignore}l<{\endignore}} 

\newcolumntype{x}[1]{%
>{\centering\hspace{0pt}}p{#1}}%

\usepackage{latexsym}
\usepackage{cite}
\usepackage{multicol}
\usepackage{eurosym}
\usepackage{xspace}
\usepackage{pdfpages}
\usepackage{verbatim}
\usepackage{stfloats}

\newenvironment{hproof}{{\noindent \it Proof:~}}

\newtheorem{problem}{Problem}
\newtheorem{proposition}{Proposition}

\newcommand{\name}{LOKO}

\DeclareMathOperator*{\argmax}{arg\,max}

\DeclareMathOperator{\Tr}{Tr}
\DeclarePairedDelimiter{\diagfences}{(}{)}
\newcommand{\diag}{\operatorname{diag}\diagfences}

\title{\name{}: Localization-aware Roll-out Planning\\for Future Mobile Networks}
\author{Antonio~Albanese,~\IEEEmembership{Student Member,~IEEE,}
Vincenzo~Sciancalepore,~\IEEEmembership{Senior Member,~IEEE,}
Albert~Banchs,~\IEEEmembership{Senior Member,~IEEE,} Xavier~Costa-P\'erez,~\IEEEmembership{Senior Member,~IEEE}
\thanks{\textit{A. Albanese is with NEC Laboratories Europe and University Carlos III of Madrid, 28911 Legan\'es, Spain. \newline
V. Sciancalepore is with NEC Laboratories Europe, 69115 Heidelberg, Germany. \newline
Xavier Costa-P\'erez is with i2cat, ICREA and NEC Laboratories Europe, 08034 Barcelona, Spain. \newline
Albert Banchs is with University Carlos III of Madrid and IMDEA Networks Institute, 28911 Legan\'es, Spain.\newline
Email of corresponding author: antonio.albanese@neclab.eu.}}%
}

\begin{document}

\IEEEtitleabstractindextext{

\begin{abstract}
The roll-out phase of the next generation of mobile networks (5G) has started and operators are required to devise deployment solutions while pursuing \emph{localization} accuracy maximization. Enabling \emph{location-based services} is expected to be a unique selling point for service providers now able to deliver critical mobile services, e.g., autonomous driving, public safety, remote operations.
In this paper, we propose a novel roll-out base station placement solution that, given a \emph{Throughput-Positioning Ratio} (TPR) target, selects the location of new-generation base stations (among \emph{available candidate sites}) such that the \emph{throughput} and \emph{localization accuracy} are jointly maximized. Moving away from the canonical position error bound (PEB) analysis, we develop a realistic framework in which each positioning measurement is affected by errors depending upon the actual wireless channel between the measuring base station and the target device. Our solution, referred to as \name{}, 
is a fast-converging algorithm 
that can be readily applied to current 5G (or future) roll-out processes. \name{} is validated by means of an exhaustive simulation campaign considering \emph{real existing deployments} of a major European network operator as well as synthetic scenarios.
\end{abstract}
\begin{IEEEkeywords}
Optimization, Semidefinite Relaxation (SDR), Roll-out, Planning, Localization, B5G, 5G.
\end{IEEEkeywords}
}

\maketitle

\section{Introduction}

The next generation of cellular networks (5G) is being bootstrapped by major network operators. Mobile operators are promptly responding to the ever-increasing demand of ubiquitous and reliable cellular communications but facing severe challenges in nurturing new business initiatives with innovative services~\cite{Ahokangas2019}. Constructing a new set of user contextual information might be the winning strategy to kick-start novel business opportunities: \emph{accurate network-based positioning} is the epitome of such advanced features. Thus, a novel, agile and efficient roll-out strategy is mandatory to support the unprecedentedly high requirements the next generation of mobile networks is expected to deliver.
However, positioning accuracy may come at an unaffordable implementation cost, hence operators need to find a viable solution to keep the Total Cost of Ownership (TCO) of the next-generation Radio Access Network (RAN) as low as possible while minimizing the impact on the existing infrastructure~\cite{Oughton2018}.

Localization accuracy is one of the widely accepted Key Performance Indicators (KPIs) for 5G services and market technology enablers. New common commercial use cases require a (horizontal) positioning error within $3$m or $10$m in indoor or outdoor scenarios, respectively, for at least $80$\% of the User Equipments (UEs) in the service area~\cite{3gpp_nr_positioning}. In addition to those general use cases, the option for \textit{higher accuracy positioning} was defined in~\cite{3gpp_tech_enabl} to guarantee UEs localization with low latency and high reliability. For instance, 5G should be able to retrieve the position of vehicles moving at a speed of (up to) $200$~km/h with less than $1$~m accuracy in more than 95 \% of service area. Finer positioning capabilities are an added-value that would benefit forthcoming network-based use-cases, e.g., tracking of goods and products within kitting processes in manufacturing, support for automated guided vehicles (AGVs) in logistic operations or victims and people in distress localization during search-and-rescue operations~\cite{albanese2020,Elayoubi2019}.

Although Global Navigation Satellite Systems (GNSSs) may provide reasonably accurate users positions, their navigation signals may incur in severe attenuation originated by e.g., debris, tree foliage, metal or wooden objects or shadowing in urban canyons thereby calling for a more reliable backup solution such as an enhanced network-based localization system. Furthermore, GNSSs are managed by third parties out of the scope of network operators, which seek increasingly higher control on such information to pursue the above-mentioned and all the upcoming business opportunities. In particular, 3rd Generation Partnership Project (3GPP) standard guidelines do not define any Application Programming Interface (API) for the purpose, thus requiring users who would willingly share their GNSS-based information to keep a client application running on their UEs, which would have a non-negligible impact on their battery life. By boosting network-based localization, network operators would benefit from the possibility of retrieving users' positions with unprecedented high accuracy without relying on UEs sharing their GNSS-based location on the data plane, therefore opening up a series of new business revenue streams for the heavily location-oriented services of the future network generations~\cite{albanese2021responders}.

The current fifth generation (5G) cellular network introduces advanced radio features, like ultra-wide bandwidth for better time resolution, additional frequency bands and massive Multi-Input-Multi-Output (MIMO) with millimeter-wave (mmWave) communications  for Angle-of-Arrival (AoA) estimation, propelling 
a new wave of networks able to directly infer positioning information with high accuracy without making use of third-party technologies~\cite{Talvitie2019}.
Indeed, 5G is expected to provide enhanced positioning capabilities, achieving (horizontal) accuracy up to $0.3$m ($99.9\%$ availability) with $15$ms latency after its complete roll-out~\cite{3gpp_discussion_localization}. From an architectural perspective, multiple radio access technologies (RATs) can be used to deliver diverse localization performances based on given requirements whereas the separation of user and control planes enables real-time positioning in the user plane by virtue of the ever-growing edge cloud~\cite{Liu2017}. Furthermore, future beyond-5G (B5G) networks will benefit from the ground-breaking Reconfigurable Intelligent Surface (RIS) technology, namely passive reflectarrays able to backscatter  and  phase-shift  the  electromagnetic waves impinging upon them by individually controlling their antenna elements~\cite{Mur21}. By doing so, RISs transform the channel into an optimization variable overcoming the traditional postulate of an uncontrollable propagation environment. Although  modern localization techniques can also take advantage of measurements from non-line-of-sight (NLoS) links, e.g., through simultaneous localization and sensing (SLAM), RISs will help circumvent line-of-sight (LoS) blockage, thus drastically improving the overall localization accuracy even in harsh propagation environments~\cite{Wymeersch2020}.

Even if such new features bring in major accuracy improvements on a single-link basis per se (e.g., leveraging measurements provided by a single Base Station (BS)), the resulting performance still depends on the base stations locations. Thus, positioning performance needs to be thoroughly addressed during the network planning phase to unleash its full potential when leveraging on measurements performed by multiple BSs. Therefore, a proper RAN deployment process becomes an impelling task: while previous RAN design solutions in the literature focus only on throughput maximization and coverage optimization, they mostly disrupt positioning accuracy performance due to the geometric dispersion of base stations or unexpected channel conditions~\cite{Schloemann2016}. Furthermore, with the advent of network slicing, network operators may shift their focus from the classical enhanced Mobile BroadBand (eMBB) use case, whose performance is mainly related to network throughput, depending on their major customers needs~\cite{Shafi2017}. For instance, a tenant from the automotive vertical sector may have lower throughput requirements in favor of higher vehicle positioning accuracy, especially in the context of the rising autonomous driving trend. In this regard, operators may take advantage of 5G (and beyond) base stations to increase the accuracy of the positioning procedures while still keeping resource allocation efficiency and end-user rates above reasonable quality levels.

In this paper, we propose a pioneering method to roll out a mobile network selecting the best base station locations among available candidate sites defined by the network operator in a pre-negotiation phase with municipal and private parties, called LOcalization-aware radio planning Keying on throughput Optimization (\name{}). Our solution is the first-of-its-kind to jointly pursue unprecedented Key Performance Indicators (KPIs), namely user positioning, throughput and fairness while considering existing network deployment limitations. In particular, differently from the canonical approach in the literature, we do not neglect the dependence of single-link positioning accuracy on the actual distance between the measuring BS and the target User Equipment (UE), thus ruling out most trivial deployment solutions, e.g. regular polygon-shaped deployments.
The contributions of our paper can be summarized as follows: we $i$) first bring the localization performance into the picture of new RAN planning, $ii$) derive a suitable formulation of the \emph{Position Error Bound (PEB)} for Time-of-Arrival (ToA) measurements accounting for the dependence of each error on the actual related measurement, $iii$) cast the roll-out process of upcoming network generation (e.g., 5G) into a novel optimization problem that jointly accounts for PEB optimization and throughput maximization, $iv$) introduce a novel parameter, dubbed as \emph{Throughput-Positioning Ratio (TPR)}, which trades off between such KPIs according to the operator requirements, $v$) devise a Block Coordinate-based algorithm that iteratively applies the Semidefinite Relaxation and Bisection techniques to different routines, $vi$) provide convergence proof of the proposed algorithm and $vii$) carry out an exhaustive simulation campaign based on synthetic and real base station deployments (from a major European network operator) to prove the effectiveness of our approach.

The remainder of this paper is as follows. Section~\ref{s:preliminaries} provides a summary of 5G-NR deployment modes and a primer on cellular-based localization; Section~\ref{s:overview} introduces the \name{} framework along with the novel concept of Throughput-Positioning Ratio (TPR), defining the desired flavour of the joint BS deployment. Section~\ref{s:model} presents the radio planning problem and proceeds by splitting it in two constituent subproblems, while Section~\ref{s:relax} describes their algorithmic solution, which is in turn numerically validated by simulations presented in Section~\ref{s:results}. Section~\ref{s:related} includes a survey of the related literature and, finally, Section~\ref{s:conclusions} concludes the paper.

\section*{Notation} 
We let $\mathbb{R}$ denote the set of real numbers. We use $\mathbb{R}_+$, $\mathbb{R}^n$, and $\mathbb{R}^{n\times m}$ to represent the sets of non-negative real numbers, $n$-dimensional real vectors, and  $m\times n$ real matrices, respectively. Vectors are denoted by default as column vectors and written in bold font. Notable examples are $\vv{1}_n$ and $\vv{1}_{n\times n}$, which respectively denote the unit vector and the unit square matrix of size $n$, while $\vv{I}_n$ is the identity matrix in $\mathbb{R}^{n\times n}$. Subscripts represent an element in a vector and superscripts elements in a sequence. For instance, $\vvi x{m} =   [ x^{(t)}_1, \dots,  x^{(t)}_n   ]^\intercal$ is a vector from $\mathbb{R}^n$ and  $x^{(m)}_i$ is its $i$th component. Operator $(\cdot)^\intercal$ represents the transpose operation, $\text{Tr}\,(\cdot)$ denotes the trace of a square matrix, $||\cdot||$ indicates the L2-norm of a vector and $\nabla$ denotes the gradient operator.  Finally, $\odot$ indicates the Hadamard or element-wise product between two matrices while $\diag{\cdot}$ returns the main diagonal of its argument.

\section{Preliminaries}
\label{s:preliminaries}
Hereafter, we detail the state-of-the-art of legacy and 5G RAN deployments thus shedding  light on recent positioning findings.
\subsection{Non-Standalone Deployment: Legacy and 5G RANs}
\label{s:lte_5g_compatibility}
Recently, 3GPP has defined the Non-Standalone (NSA) deployment mode as a 5G-NR deployment option that relies on the existing LTE core network (EPC) while applying 5G-NR features to user plane functions only~\cite{3gpp_rel16}. More and more operators prefer it in the short-term: they may ease the complexity of deploying a nationwide Standalone (SA) 5G network from scratch (as opposed to NSA) by performing only a selective deployment of 5G-NR base stations (gNBs) with no need of a proper 5G Core (5GC). 
Therefore, we consider hybrid scenarios with legacy\,\footnote{Our main findings may be easily applied to 5G hybrid scenarios employing any legacy radio access technology, including LTE, 3G and so on. For the sake of simplicity, in this paper we only model the LTE technology.} base stations already in place. Please note that our solution may be readily applied to SA deployment mode, as 5G-only networks are a subset of such hybrid scenarios. 

\textbf{Hybrid Deployment.} To guarantee backwards compatibility in the new RAN design, 
low frequencies---employed by LTE---will continue being used to provide sufficient signal level in rural, suburban and urban areas~\cite{Wan2019}, while most network operators have favored the higher C-Band for their first 5G-NR deployments with up to five-time channel bandwidth increase compared to LTE~\cite{5g_observatory}. %
Indeed, a higher bandwidth associated with a higher spectral efficiency allows gNBs to offer much larger throughput w.r.t. a classical eNB. Additionally, a higher bandwidth allows for a proportionally higher resolution in the delay domain, making multipath easier to address, thus alleviating the measurement error in range-based positioning, see e.g. the wide-spread usage of Ultra Wide Bandwidth (UWB) signals for localization~\cite{nec_nsdi_19}.

\subsection{Time-based positioning}
\label{s:ToA}

Time-based positioning is a fundamental positioning technique, which combines Time-of-Flight (ToF) measurements related to a set of (at least three) base stations to derive the location of some User Equipment (UE) by means of the multilateration technique. 
To this aim, the UE receives periodic Positioning Reference Signals (PRSs) from the serving and the adjacent base stations and records their respective Time of Arrivals (ToAs)\footnote{For the sake of tractability, we focus our analysis on ToA only. It is worth pointing out that Time-Difference-of-Arrival (TDoA) eliminates the need for knowing the absolute transmission times of PRSs by subtracting the ToAs of a reference base station from the collection of measurements. However, ToA and TDoA provide the same theoretical lower bound in conventional scenarios. Besides, the PEB in NLoS conditions with some a-priori knowledge on the biases is lower-bounded by the PEB attained in perfect LoS conditions. We refer the reader to~\cite{Kaunel2015, ToA_NLoS} for further details.}. \change{PRSs are sent within Positioning subframes, which are designed as Low Interference Subframes (LISs), i.e. with no concurrent data transmission over the Physical Downlink Shared CHannel (PDSCH). Therefore, in perfectly synchronized networks, PRSs are isolated from data transmission and are only exposed to the interference generated by other cell PRSs (with the same pattern index). Dedicating time and bandwidth resources (in terms of Physical Resource Blocks (PRBs)) to PRSs has an impact on the achievable network throughput, which is usually lower than $1\%$ in normal operating conditions, as by 3GPP recommendations~\cite{fischer2014observed}. Therefore, in our analysis, we neglect such effect as it would tie the derivation to the specific implementation of the multilateration technique}\footnote{\change{Nonetheless, our problem formulation may be extended to account for PRSs overhead by suitably scaling down the achievable throughput expression in the problem formulation.}}. Below we derive the lower bound on the positioning accuracy provided by a technique of this kind, which in the rest of the paper we employ as positioning performance metric. 

\textbf{Time-of-Arrival measurements.} Let us consider a cellular network consisting of $B$ base stations (BSs) deployed at coordinates $\vv{q}_i = (x_i,y_i)^\intercal$ and one UE performing ranging measurements\footnote{We recognize that in practice several options are available to improve the performance of a localization technique solely based on time measurements, e.g. MIMO systems for estimating Angle-of-Arrival (AoA) and Angle-of-Departure (AoD), cooperation among base stations, and so forth. However, in this paper we propose the first localization-aware deployment solution and show the implications of such aspects on real deployment snapshots. Therefore, we argue that the inclusion of the above-mentioned positioning techniques is out of scope.} from all available BSs in order to infer its own geographical coordinates $\vv{p} = (x,y)^\intercal$. Let us denote the set of ToA range measurements as:
\begin{align}
    \tilde{\varrho}_i(\vv{p}) = d_i(\vv{p}) + l_i + n_{i}, \quad \forall i \in \mathcal{B},
\end{align}
where $d_i(\vv{p}) = ||\vv{p}-\vv{q}_i||$ is the true Euclidean distance between the UE and $i$th base station, $l_i$ is a positive bias modeling the excess path length due to Non-Line-of-Sight (NLoS) propagation conditions, and $n_i$ is a random error.

Following~\cite{Jourdan2008}, we assume that biases $l_i$ are uniformly distributed between $0$ and some $\lambda_i$\,\footnote{Typically, the only a-priori information about the environment is the maximum bias $\lambda_i$ while a detailed description of its distribution may not be available.}. This assumption strives to minimize the burden on network operators performing an extensive measurement campaign. Indeed, estimating the statistics of the ToA measurement biases in the area for every possible BSs deployment is a particularly demanding task, which, in turn, leads to a difficult-to-manage PEB expression in the light of its optimization. Therefore, we limit the measurement effort to the maximum bias values and assume their statistics to be uniform between $0$ and such values, i.e. $\lambda_i$'s. Measurement errors $n_i$ are modeled as independent Gaussian random noises with zero mean and variance $\sigma^2_i$ depending on the distance $d_i$~\cite{variance_distance}, namely $\sigma^2_i = \sigma^2_0 (d_i/d_0)^{\alpha_i}$, where $\alpha_i \geq 0$ is the path-loss exponent and $\sigma^2_0$ is the variance at a reference distance $d_0$. 
Indeed, the higher the distance, the higher the number of indirect propagation paths, thereby exacerbating the task of correctly estimating the time of arrival of the dominant path at the UE side. Moreover, this preservers the generality of our model as it includes the distance-invariant one by setting $\alpha_i = 0$. 

The probability density function (pdf) of unbiased range measurements $\varrho_i$ can be written as
\begin{equation}
    f_{\varrho_i}(\varrho_i|\vv{p})\! = \! \frac{1}{\lambda_i}\!\left[Q \! \left( \frac{\varrho_i \! - \! d_i(\vv{p})\! - \!\lambda_i}{\sigma_i}\right)\! - \!Q\! \left(\frac{\varrho_i - d_i(\vv{p})}{\sigma_i}\right)\! \right],
    \label{eq:ToA_pdf}
\end{equation}
where $Q(z) = \frac{1}{\sqrt{2\pi}} \int_z^{+\infty} e^{-t^2} dt$ is the Gaussian Q-function\footnote{\change{A closed-form expression for the TDoA case is not available. However, numerical results show that the lower the lumped error $l + n$ on the reference ToA compared to the lumped error on the current ToA, the better Eq.~\eqref{eq:ToA_pdf} approximates the pdf of the resulting TDoA measurements.}}.

\textbf{Position Error Bound.} To assess the localization performance of cellular networks, we leverage on the \emph{Cram{\'e}r-Rao Lower Bound} (CRLB), which provides a lower bound on the covariance matrix of any unbiased estimator. This makes our problem completely general, without tailoring it to a specific localization technique (e.g., \cite{Yu2019}), while still providing useful insights on the expected performance of a positioning system deployed in the field~\cite{CRB}. The choice of assessing localization performances by means of the PEB stems from the need for a general and reasonably compact expression to optimize the localization capabilities of the network, which is independent of the specific implemented localization solution. This approach goes along the lines of making use of the capacity definition in place of the sum-rate, which in reality would be affected by, e.g., the specific scheduling and resource allocation at the BSs.    
Let us define the estimated UE position as $\hat{\vv{p}} = (\hat{x},\hat{y})^\intercal$ based on range measurements vector $\vv{\varrho} = (\varrho_1,\dots, \varrho_B)^\intercal$. The covariance matrix of $\hat{\vv{p}}$ satisfies:
\begin{equation}
    \textup{cov}(\vv{\varrho}) = \mathbb{E}_{\vv{\varrho}}\left[(\vv{p} - \hat{\vv{p}})(\vv{p} - \hat{\vv{p}})^\intercal\right] \succeq \vv{J}^{-1}_{p},
    \label{eq:covariance_semipositive}
\end{equation}
where $\mathbb{E}_{\vv{\varrho}}[\cdot]$ represents the expectation over $\vv{\varrho}$, $\succeq$ indicates that $\textup{cov}(\vv{\varrho}) - \vv{J}^{-1}_{\vv{p}}$ is positive semidefinite and $\vv{J}_p$ is the Fisher Information Matrix (FIM) for a given $\vv{p}$ and base station deployment $\vv{q} = (\vv{q}_1,\dots,\vv{q}_B)^\intercal$, defined as:
\begin{equation}
    \vv{J}_p = \mathbb{E}_{\vv{\varrho}}\left[\left(\nabla_{\vv{p}} \ln f(\vv{\varrho}|\vv{p})\right) \left(\nabla_{\vv{p}} \ln f(\vv{\varrho}|\vv{p})\right)^\intercal \right].
    \label{eq:FIM}
\end{equation}
Note that, since range measurements are assumed to be independent, we have  $f(\vv{\varrho}|\vv{p}) = \prod_{i=1}^B f_i(\varrho_i|\vv{p})$. From Eq.~\eqref{eq:covariance_semipositive}, it yields the following 
\begin{equation}
    \sqrt{\mathbb{E}_{\vv{\varrho}}\left[(x - \hat{x})^2 + (y - \hat{y})^2\right]\vphantom{\Tr(\vv{J}^{-1}_{p})}} \geq \sqrt{\Tr(\vv{J}^{-1}_{p})}.
\end{equation}
The value $\sqrt{\Tr(\vv{J}^{-1}_{p})}$ is dubbed as \emph{Position Error Bound} (PEB) and represents a lower bound on the mean-square error (MSE) of the distance between the UE position and its corresponding estimate.
The PEB of ToA measurements with uniformly distributed measurement biases can be written as the following (\!\cite{Jourdan2008})\footnote{This definition is not limited to UWB, but it applies to any ToA-based localization system regardless of its transmit power, operating bandwidth and propagation environment.}
\begin{equation}
    \beta(\vv{p}) = \left(\frac{\sum\limits_{i \in \mathcal{B}} \nu_i}{\sum\limits_{\substack{i<j, \\ i,j \in \mathcal{B}}} \nu_i \nu_j \sin^2(\theta_j - \theta_i)}\right)^{\frac{1}{2}},
    \label{eq:PEB}
\end{equation}
where $\nu_i$ is a function of $d_i$ and $\lambda_i$:
\begin{equation}
    \nu_i(d_i,\lambda_i) = \frac{1}{\lambda_i \sigma_i(d_i) \pi \sqrt{2}} \int_{-\infty}^{+\infty} h(y,\lambda_i,d_i) \, dy,
    \label{eq:weights}
\end{equation}
with $d_i = d_i(\vv{p})$, $\lambda_i = \lambda_i(\vv{p})$, $\theta_i = \theta_i(\vv{p})$ the angle between $\vv{p}$ and $\vv{q}_i$ w.r.t. the horizontal axis, and $h(y,\lambda,d) =$
\begin{align}
    \frac{\left(\!e^{-\left(y + \frac{\lambda}{\sigma(d) \sqrt{2}}\right)^2}\!\!\!\left( 1 + \frac{\alpha \lambda}{2d} + \frac{\alpha \sigma(d)}{d\sqrt(2)}y\right) \!- \!e^{-y^2}\!\!\left( 1 + \frac{\alpha \sigma(d)}{d \sqrt{2}}y\right)\right)^2}{Q\left(\sqrt{2}y\right) - Q\left(\sqrt{2}y + \frac{\lambda}{\sigma(d)}\!\!\right)}.
    \label{eq:h}
\end{align}

We would like to point out that accounting for the dependence of the measurement errors on the actual distance is key for making the PEB a reasonable metric for our deployment solution. Indeed, under the assumption of independent errors, namely for $\alpha_i = 0, \, \forall i \in \mathcal{B}$, it is well-known that the lowest value of PEB occurs at the center of a regular $B$-sided polygon of any side length~\cite{Levanon2000}. Therefore, one trivial solution to the PEB optimization problem would be to spread the BSs regularly around the deployment area far enough from each UE to allow approximating their position with the polygon center. On the contrary, if single-link errors depend on the actual distance, this kind of solution is ruled out and some other deployment schemes not necessarily regular around each UE might be possible, as shown in the following.

The PEB in Eq.~\eqref{eq:PEB} depends on a twofold factor: $i)$ the propagation conditions experienced by the transmitted signals through weights $\nu_i$ and $ii)$ the angular positions of each base station (BS) w.r.t. the UE. 
In realistic deployments, the performance gain expected from 5G-NR with respect to the previous generation is difficult to achieve only by resorting on physical layer improvements but rather requires a paradigm shift in the RAN roll-out by considering the positioning performance in Eq.~\eqref{eq:PEB} as a target KPI in the planning phase.

\begin{figure*}[t]
      \centering
      \includegraphics[ width=.77\linewidth ]{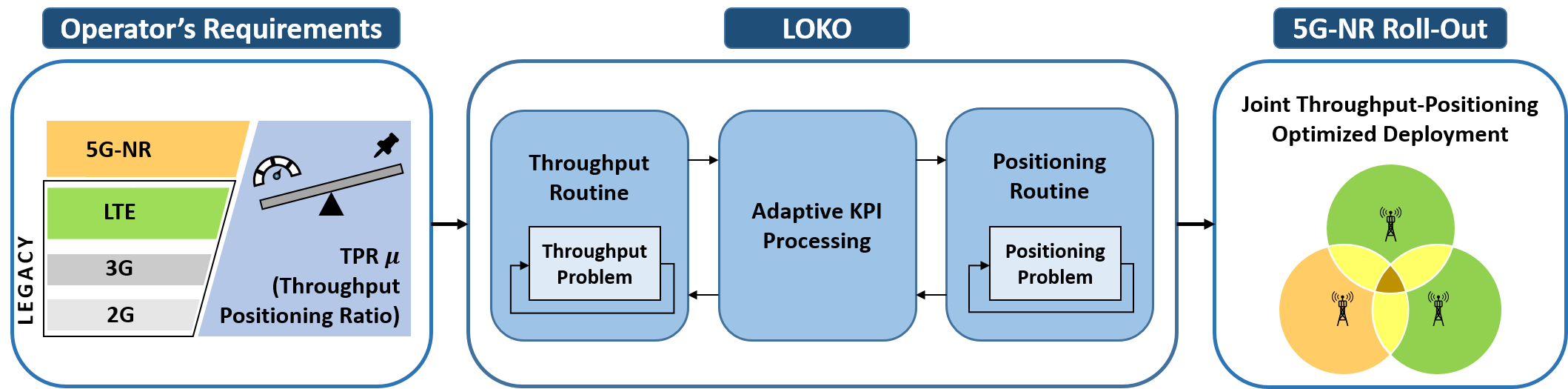}
      \caption{Overview of \name{}'s building blocks}
      \label{fig:system_model}
\end{figure*}

\subsection{Measurement impairments and randomness} 
As described in Section~\ref{s:ToA}, NLoS conditions introduce a bias in the range measurements due to the blockage of the Line-of-Sight (LoS) direct path with the serving base station, e.g. in presence of shadowing. In such case, the UE measures a positively biased distance since the dominant received signal propagates along an indirect path, which is longer than the obstructed direct one. Travelling a longer path, the signal experiences a higher attenuation that is usually modeled according to a log-normal distribution with $\sigma_s$ standard deviation. 

It is worth pointing out that the direct path may still be received by the UE if the additional attenuation introduced by the blocking obstacles is not enough to lower the received power below the UE sensitivity. Nevertheless, the refraction experienced by such path propagating through the obstacles leads to an additional delay, whose value depends on the refractive index of the obstacles material. Therefore, there is a close proportional link between the bias distribution in the range measurement process and the magnitude of the wireless channel shadowing, namely between $\lambda$ and $\sigma_s$, which we take into account in our numerical evaluation in Section~\ref{s:results}. 

\section{\name{} Framework overview}
\label{s:overview}

A proper radio planning lies in selecting BSs locations and configuration parameters in order to provide full coverage and meet the required Quality of Service (QoS) in a target area, while keeping the network operator CAPEX and OPEX as low as possible. State-of-the-art solutions including commercial radio planning software packages provide heuristic solutions of the problem in its classical flavour, which takes no account of users' localization. With \name{}, we take a step forward by designing an adaptive technique to jointly optimize network throughput, user fairness and positioning performances.

\textbf{Network integration.}
Operators willing to incorporate 5G-NR in their live network likely have got an up-and-running legacy deployment, such as LTE or 3G. The most convenient way to bootstrap 5G-NR deployment is to employ the NSA option, as discussed in Section~\ref{s:lte_5g_compatibility}. Therefore, \name{} takes the pre-existing LTE network into account and improves upon its throughput and positioning performances, which represent the guaranteed baseline.       

\textbf{Deployment settings.} In a realistic operation scenario, network operators must not exceed their budget, namely they can afford only a limited given number of new base stations, while meeting logistical and administrative constraints during the deployment phase. On the one hand, they are required to comply with municipal restrictions in terms of electromagnetic emissions and public land occupation. On the other hand, they seek one-to-one agreements with private parties to install network equipment in their facilities. In addition to the administrative regulations, there are also physical constraints to deal with. For instance, when deploying a base station on the roof of a building, there might be only a specific location on the roof where network operators could physically place the base station. Likewise, not all buildings in a city may allow a base station on their roof. Thus, \name{} takes as input the set of candidate sites that satisfy the above-mentioned requirements, and selects the corresponding subset wherein new base stations will be deployed without violating the budget constraint. We would like to underline that there is no intrinsic limitation on the set of candidate sites, which could possibly include any location in the service area, e.g. all buildings in a city. Therefore, considering a set of candidate sites does not introduce any loss of generality in our proposed solution. Note that this set contains also the sites with existing LTE BSs to account for the possibility of co-locating LTE and 5G BSs and thus reducing installation costs; however, there is no guarantee that the set of existing LTE locations would satisfy the positioning accuracy requirements of the operator in the 5G era. We propose to sample the target area by means of $T$ test points, wherein a typical UE should meet the desired requirements in terms of throughput and positioning performances. The test point distribution should match the expected distribution of cellular users in such area. Nevertheless, our solution is not tailored to a specific set of test points and can be readily applied to any user distribution.

\begin{table}[t]
\caption{Terminology}
\label{tab:terminology}
\centering
\resizebox{\linewidth}{!}{%
\begin{tabular}{cc|cc}
\textbf{Acronym} & \textbf{Parameter} & \textbf{Acronym} & \textbf{Parameter}\\  
\hline
\rowcolor[HTML]{EFEFEF}
RAT & Radio Access Technology & RAN  & Radio Access Network \\
ToA & Time-of-Arrival & GNSS & Global Navigation Satellite System \\
\rowcolor[HTML]{EFEFEF}
CRLB & Cram\'er-Rao Lower Bound  & PEB            & Position Error Bound\\
BS        & Base Station    & eNB/gNB  & LTE/5G Base Station \\
\rowcolor[HTML]{EFEFEF}
UE & User Equipment & CS & Candidate site\\
TPR & Throughput-Positioning Ratio & SDR & Semidefinite Relaxation\\
\rowcolor[HTML]{EFEFEF}
BCA/D & Block Coordinate Ascend/Descend & FP & Fractional Programming 
\end{tabular}%
}
\end{table}

\textbf{Flavour tuning.} Throughput and localization performances are different metrics that employ different units of measurements. As there is a trade-off between throughput and location accuracy, this raises the need for a tuning parameter $\mu$, dubbed as Throughput-Positioning Ratio (TPR), which reflects the desired flavour of the planning solution. Indeed, TPR $\mu$ takes into account how much an operator values a unit gain towards throughput or localization performances, in other words how many $bit/s$ would it put at stake to improve the PEB by $1$ meter. 

\textbf{Iterative Solution.} As shown in~Fig.~\ref{fig:system_model}, \name{} takes as input the above-mentioned sites and the desired planning flavour and performs an iterative procedure, which quickly converges to a planning solution fulfilling the QoS requirements of the operator.

Lastly, Table~\ref{tab:terminology} lists the terminology used throughout the manuscript for better readability.

\section{System Model}
\label{s:model}

We formalize the overall problem in order to strike the perfect balance between throughput and localization performances experienced by a typical user in the deployment area by means of a tunable parameter, namely Throughput Positioning Ratio (TPR). Specifically, we first formulate it as a joint optimization problem and discuss its tractability. Then, we boil down its complexity by considering two constituent problems and manipulating them to obtain algorithmic sub-optimal solutions, which are described in detail in Sections~\ref{s:throughput_solution} and~\ref{s:positioning_solution} respectively. 

\subsection{Joint Radio Planning Problem}\label{s:joint_problem}
As previously discussed in Section~\ref{s:overview}, our target area encompasses pre-deployed legacy base stations (in our example, LTE) providing the performance baseline that we aim to improve by means of our solution, namely \name{}. 

Let us characterize the set of base stations $\mathcal{B}$ as the union of two disjoint subsets (or tiers) $\mathcal{E}$ and $\mathcal{G}$ of LTE (eNBs) and 5G-NR (gNBs) base stations, respectively. The former encloses all already deployed eNBs while the latter includes all available gNBs that can be deployed without exceeding the allocated budget by the network provider. Note that we assume full frequency reuse, and same transmission parameters and propagation conditions within a single tier\footnote{These assumptions serve the purpose of a more compact notation, although our analysis may be readily extended.}, while considering no interference between the two, as per Section \ref{s:lte_5g_compatibility}. Therefore, we respectively denote by $P_k$, $f_k$, $W_k$, $\alpha_k$ and $\sigma_{s_{k}}$ the transmit power, central frequency, bandwidth, propagation factor and shadowing standard deviation associated to tier $k$, where $k$ can exclusively take values $e$ or $g$ according to the corresponding tier $\mathcal{E}$ or $\mathcal{G}$, respectively. We seek a \emph{hybrid user association strategy}, namely a typical UE associates to the base station showing the highest downlink Signal-to-Interference-plus-Noise-Ratio (SINR) among the available ones, given that it allows satisfying the positioning accuracy requirements\footnote{As per a conventional maximum-SINR association strategy, the UE prefers to associate to a gNB if another eNB offers the same SINR. When two or more gNBs are eligible for association, the UE performs a random decision.}.

Let us write the average downlink SINR experienced in the slow-fading regime by a typical UE at coordinates $\vv{p}$ associated to base station $i$ belonging to tier $k$ as
\begin{equation}
\gamma_i(\vv{p}) = \frac{P_k g_i(\vv{p})}{P_k g'_i(\vv{p}) + N_0W_k},
\end{equation}
where $g_i(\vv{p}) = (4 \pi f_k d_i(\vv{p})/v)^{-\alpha} \exp{(-\sigma_{s_k}^2/2\xi^2 )}$, with $\xi = 10/\log(10)$ and $v$ indicating the speed of light in vacuum, is the average channel attenuation in presence of log-normal shadowing, $N_0$ is the two-sided noise power spectral density and $g'_i(\vv{p})$ denotes the power-normalized interference experienced by the UE, namely $g'_i(\vv{p}) = \sum_{n \in \mathcal{K} \setminus \{i\}} g_n(\vv{p})$, where $\mathcal{K} = \{\mathcal{E},\mathcal{G}\}$
indicates which subset (5G or LTE) base station $i$ belongs to. Note that interference arises only from active base stations belonging to the same tier as the serving one, as previously discussed in Section~\ref{s:lte_5g_compatibility}. Beamforming capabilities of MIMO systems could reduce the interference between neighboring base stations operating on the same bandwidth. However, especially in dense-urban scenarios, UEs may be spatially spread around each BS, thereby leading to wide beams at the limit corresponding to a Single-Input-Single-Output (SISO) radio configuration.

\textbf{Decision variables.}
To facilitate the installation procedure, we assume that base stations may be deployed only at operator-defined candidate sites (CSs), thus calling for coordinate discretisation. It is worth pointing out that \name{} may likewise be executed even if there is not any a-priori knowledge of CSs by taking as input any sampling of the target area. We name $\mathcal{S}$ the set of such CSs with coordinates $\vv{s}_j \in \Omega$, $\forall j \in \mathcal{S}$, where $\Omega \in \mathbb{R}^2$ is the target deployment area, being $S$ the cardinality of $\mathcal{S}$. For example, $\mathcal{S}$ may be a superset of the sites where LTE base stations have already been deployed. Indeed, eNBs coordinates are fed into the problem, and kept fixed during the optimization process. 

Our planning solution outputs the set of CSs wherein gNBs will be deployed, while providing the association between the typical UE in each test point and its optimal base station under the assumption of a single UE per test point. Therefore, we introduce decision variables $x_j \in \{0,1\}$ and $a_{tj} \in \{0,1\}$ indicating whether a gNB is deployed at CS $j$ and the association between a typical UE at test point $t$ and CS $j$, respectively. 

\textbf{Network performances.} In principle, to calculate the network throughput experienced at each test point, we should average the user throughput over time. However, user rate depends on the serving base station scheduler, which assigns the optimal Modulation and Coding Scheme (MCS) given the current channel quality. 
Indeed, for the sake of tractability, we only consider Shannon's channel capacity $W_k\log_2(1+\gamma_i(\vv{p}_t))$, assuming that the base station schedulers perform near-optimal MCS assignments and the entire bandwidth of the best station is assigned to the test user. It provides an upper bound on the throughput that a typical UE experiences at test point $t$ with coordinates $\vv{p}_t \in \Omega$ while being served by base station $i$. As far as positioning performances go, we refer to the PEB defined in Section~\ref{s:ToA}\,\footnote{We only refer to Single-Input-Single-Output (SISO) Capacity. Although 5G-NR has MIMO capabilities, we spare the additional model complexity as it would provide analogous results at least with quasi-uniform user distributions.}. By using this formulation, we depart from the conventional assumption that each measurement error is independent of the actual measurement thereby accounting for the additional attenuation and fading experienced by UEs that are further away from the measuring BS, thus improving the validity of our system model. Said that a typical UE receives PRSs from active BSs, we need to discriminate only between LTE and 5G network association, relaxing the need of finely optimizing association variables $a_{tj}$.

\textbf{Throughput Positioning Ratio (TPR $\mu$).} To properly account for heterogeneous objectives, we need to introduce in the objective function a scalar factor $\mu$ (TPR) that trades off between throughput and positioning performances. In the extreme case of $\mu=0$, our optimization will provide only throughput guarantees. Conversely, when $\mu \rightarrow \infty$, the solution will optimize positioning performances regardless of the experienced throughput values. A detailed evaluation of such a parameter is provided in Section~\ref{s:real_depl} and Table~\ref{tab:tpr}.

\textbf{Problem formulation.} We formulate the joint throughput-positioning planning problem\footnote{The overall deployment cost or power consumption vary according to the selected sites. Although we do not explicitly account for them, the model can be easily extended by adding two corresponding linear budget constraints, with little to no modifications to the inner loops described in Sections~\ref{s:throughput_solution} and~\ref{s:positioning_solution}.}. It is worth pointing out that our joint goal is to maximize the fairness among test points, both throughput and positioning-wise. Note that this objective maps onto a max-min problem as follows
\begin{problem}[Joint Throughput-Positioning Radio Planning]\label{problem:joint_optimization}
\begin{subequations}
{\begin{align}
   & \max_{\vv{A},\vv{x}} && \min_{t \in \mathcal{T}} \,(\vv{r} - \mu \vv{b}) \label{eq:joint_objective}\\
   & \textup{subject to} && \vv{A} \leq \vv{1}_T \,\vv{x}^\intercal,    \label{eq:constraint_a_leq_x}\\
   & && \vv{A} \, \vv{1}_S \leq \vv{1}_T,  \label{eq:constraint_sum_a_leq_1}\\
   & && \vv{x}^\intercal  \vv{1}_S  \leq G, \label{eq:constraint_cap}\\
   & && \vv{x} \in \{0,1\}^S,\quad\vv{A} \in \{0,1\}^{T \times S},  \label{eq:constraint_binary_x_a}
\end{align}}
\end{subequations}
\end{problem}
wherein $\vv{A}$, with $\{\vv{A}\}_{tj} = a_{tj}$, is the association decision matrix and $\vv{x} = [x_1,\dots,x_S]^\intercal$ is the deployment decision vector. Constraint \eqref{eq:constraint_a_leq_x} states that a gNB must be deployed at a certain CS as long as one test point is associated to it, while constraint \eqref{eq:constraint_sum_a_leq_1} reflects the fact that each test point can be associated to at most one CS (thus base station), as prescribed by 3GPP. Moreover, constraint~\eqref{eq:constraint_cap} enforces the budget constraint by limiting the maximum number of deployed gNBs to $G$, where $G = |\mathcal{G}|$ is the cardinality of $\mathcal{G}$. Objective function~\eqref{eq:joint_objective} counts two contributions, namely $\vv{r}$ and $\vv{b}$, which are shortenings for $\vv{r}(\vv{A},\vv{x})$ and $\vv{b}(\vv{A},\vv{x})$, respectively. The former considers the throughput performance and its elements can be written as
\begin{align}
r_t = \sum_{j \in \mathcal{S}} \left( a_{tj}  \log_2  \left(1 + \frac{ g_{tj}}{g'_{tj} + N'}\right)\right)W_g +  a'_t c_t,
\label{eq:objective_throughput}
\end{align}
where $g_{tj} = g_j(\vv{p}_t)$, $g'_{tj} = \sum_{n \in \mathcal{S}\setminus \{j\}} g_{tn} x_n$ indicates the normalized interference experienced by a typical UE at site $t$ given deployment decision $\vv{x}$, $N' = N_0 W_g/P_g$ is the normalized noise power, $a'_t = 1 - \sum_j a_{tj}$, and $c_t$ is the throughput provided at $\vv{p}_t$ by the pre-deployed eNBs, namely
\begin{equation}
    c_t =  \max_{m \in \mathcal{E}} W_e\, \log_2  \left(1 + \gamma_m(\vv{p}_t)\right).
\end{equation}
Note that, whenever a UE is better off connecting to the LTE network, there is no need in our solution to keep track of the specific eNB, as the best association is defined beforehand by the legacy deployment.

The second contribution $\vv{b}$ takes into account the positioning performance. Its elements are of the form 
\begin{align}
b_t = (1 - a'_t)\left( \frac{\vv{\nu}_t^\intercal \vv{x}}{\vv{x}^\intercal \vv{F}_t^\intercal\vv{x}}\right) + a'_t \, u^2_t, \label{eq:objective_peb}
\end{align}
where $\vv{\nu}_t = [\nu_{1t},\dots,\nu_{St}]^\intercal$, with $\nu_{jt} = \nu_j(\vv{p}_t)$, $\vv{F}_t$ is a strictly upper triangular matrix so as $\{\vv{F}_t\}_{i,j} = \nu_{it} \nu_{jt} \sin^2(\theta_{jt} - \theta_{it})\, \forall i<j$, with $\theta_{jt} = \theta_j(\vv{p}_t)$ and $u_t$ is the PEB given by the LTE deployment as per Eq.~\eqref{eq:PEB}. 

\textbf{Analytical tractability.}
Even reduced versions of Problem~\ref{problem:joint_optimization} are NP-Hard, thus making it difficult to find its exact solution. For instance, let us consider a scenario where all users experience the same SINR towards all BSs and optimize for throughput only. In this case, the problem is equivalent to a bin packing problem, which is known to be NP-hard. Therefore, we develop a heuristic solution by means of convex relaxation techniques.

Binary constraint \eqref{eq:constraint_binary_x_a} makes Problem~\ref{problem:joint_optimization} combinatorial. Although after relaxing this constraint the set of constraints is convex, the joint objective function is still non-convex, being the sum of non-convex components. Each component calls for a different solution methodology, which makes it impractical to tackle the joint optimization problem directly. In the following, we break down the original problem into two problems and derive a sub-optimal solution for each of them. Such solutions constitute the buildings blocks of \name{} as shown in Fig.~\ref{fig:system_model}, which is eventually able to obtain a solution for the original joint optimization problem.


\subsection{Problem separation}
\label{s:throughput}
3GPP standards have heretofore favored throughput over other performance figures, making it the de-facto most compelling parameter for any cellular network deployment. However, more recently we are witnessing a reversal of the trend with the ongoing standardization of 3GPP fifth generation. In particular, 5G promises to improve several neglected figures such as end-to-end latency, support for massive IoT and enhanced positioning accuracy. 

\textbf{Throughput Problem.} Nevertheless, squeezing the most throughput out of a network deployment remains a top priority for operators, as its achievement leads to some significant reduction of CAPEX and OPEX, thus eventually increasing the operating revenues. Let us boil down Problem~\ref{problem:joint_optimization} to two constituent fundamental problems, namely Throughput Problem and Positioning Problem. First, we aim at maximizing the throughput fairness among test points while ensuring that localization performances are below a given threshold $\zeta_b$. This threshold is a key parameter of \name{} and is thoroughly analysed in Section~\ref{s:flavour}.
The above problem can be cast into the following 
\begin{problem}[Throughput Problem]\label{problem:throughput_optimization}
\begin{subequations}
{\begin{align}
   & \max_{\vv{A},\vv{x}} && \min_{t \in \mathcal{T}} \vv{r} & \label{eq:objective_throughput_vector}\\
   & \textup{subject to} && \vv{b} \leq \zeta^2_b \vv{1}_T, & \label{eq:constraint_b_leq_th}\\
   & && (\ref{eq:constraint_a_leq_x}),(\ref{eq:constraint_sum_a_leq_1}),(\ref{eq:constraint_cap}),(\ref{eq:constraint_binary_x_a}).
\end{align}}
\end{subequations}
\end{problem}

Note that $\zeta_b \leq u_t$ $\forall t$, as LTE positioning performances represent a baseline for our solution.  

\textbf{Positioning problem.} Conversely to the previous problem, one operator may optimize the positioning performances of the network while satisfying some throughput performance threshold $\zeta_r$. Again, we take the LTE network performances as baseline by assuming $\zeta_r \geq r_t$ $\forall t$. Therefore, we can define the positioning problem as follows 
\begin{problem}[Positioning Problem]\label{problem:positioning_optimization}
\begin{subequations}
{\begin{align}
  & \min_{\vv{A},\vv{x}} && \max_{t \in \mathcal{T}} \vv{b} & \label{eq:objective_peb_vector}\\
  & \textup{subject to} && \vv{r} \geq \zeta_r\vv{1}_T, & \label{eq:constraint_r_geq_th}\\
  & && (\ref{eq:constraint_a_leq_x}),(\ref{eq:constraint_sum_a_leq_1}),(\ref{eq:constraint_cap}),(\ref{eq:constraint_binary_x_a}).
\end{align}}
\end{subequations}
\end{problem}
The two problems are non-convex, thus hard to solve as they cannot be tackled with convex optimization techniques. We discuss their non-convexity property in the next section while developing suitable algorithmic solutions.

\section{Convex relaxation and solution}
\label{s:relax}
In order to provide a practical and affordable solution, we apply a convex relaxation to the above-described problems.

\subsection{Convexity Analysis} \label{s:convexity_analysis}
We first prove that Problems~\ref{problem:throughput_optimization} and \ref{problem:positioning_optimization} are non-convex. Since $\vv{r}$ and $\vv{b}$ swap roles in the problems being objective function and constraint, respectively, we can focus on one single problem, i.e., Problem~\ref{problem:throughput_optimization}, leaving the non-convexity property of Problem~\ref{problem:positioning_optimization} as an analogous proof. 

Let us consider one element of $\vv{r}$, as per Eq.~\eqref{eq:objective_throughput}. The sum of ratios (of $g'_{tj}$ that is function of $x_j$) shows the non-convexity in the optimization variable $\vv{x}$. 
%
Let us now consider the set of constraints. While constraints \eqref{eq:constraint_a_leq_x}, \eqref{eq:constraint_sum_a_leq_1} and~\eqref{eq:constraint_cap} are linear (and convex), constraints \eqref{eq:constraint_b_leq_th} and \eqref{eq:constraint_binary_x_a} are not. With regard to constraint \eqref{eq:constraint_b_leq_th}, $b_t$ provides a linear term $a'_t u^2_t$, whereas the remaining one is a product of a linear function of $a_{tj}$ and a non-convex function of $x_j$, thus making the constraint non-convex. Finally, constraint~\eqref{eq:constraint_binary_x_a} is non-convex as it enforces the binary nature of the decision variables.
First off, we relax this constraint making $a_{tj}$ and $x_j$ continuous in the closed convex set $[0,1]$. Then, we introduce a number of convex relation techniques and optimization methods tailored to address each non-convexity issue of the above-mentioned problems, and provide practical solutions. 

\subsection{Throughput Routine}\label{s:throughput_solution}
We describe the devised routine used to solve Problem~\ref{problem:throughput_optimization}. It requires joint gNB associations and deployment decisions, while taking into account the throughput provided by the pre-existing LTE deployment. Note that one gNB is deployed at CS $j$ if there exists a test point $t$ that would get higher throughput w.r.t. the LTE network, given the optimal deployment decisions at the other sites, the maximized fairness among test points and the satisfaction of the positioning constraint. Besides, the total number of deployed gNBs must not exceed the allocated budget.

Let us consider the continuous relaxation of the problem, as already mentioned in Section~\ref{s:convexity_analysis}. In the following, we develop a double-nested Block Coordinate Ascend (BCA) algorithm~\cite{Grippo2000} as detailed in Algorithm~\ref{alg:throughput_optimization}, 
wherein association matrix $\vv{A}$ is optimized while keeping decision variable $\vv{x}$ fixed and vice-versa, in an iterative fashion\footnote{To avoid notation clutter, we annotate variables computed within the inner and outer loop with $\bar{\cdot}$ and $\hat{\cdot}$, respectively.} whose convergence proof is provided in the following. 

  \begin{algorithm}[t]
    \DontPrintSemicolon
    Initialize $\vv{A},\vv{x}, \vv{y}$ to feasible values $\hat{\vv{A}}^{(k=0)},\bar{\vv{x}}^{(k=0)}$\;
    \Repeat(\hfill\emph{Outer BCA loop (over $k$)}){convergence of objective function Eq.~\eqref{eq:objective_throughput_vector} in Problem~\ref{problem:throughput_optimization}}{
        \Repeat(\hfill\emph{Inner BCA loop (over $i$)}){convergence of $\vv{r}'$ as per Eq.~\eqref{eq:objective_throughput_FP}}{
            Derive $\bar{\vv{y}}^{(i)}$ by solving Problem~\ref{problem:throughput_optimization} as $\max\limits_{\vv{y}}\,\,\min\limits_{t\in\mathcal{T}}\vv{r}'',$ \vspace{-1.5mm}\newline where $\vv{r}''$ is obtained from Eq.~\eqref{eq:objective_throughput_quadratic_transform}\;  
            Derive $\bar{\vv{X}}^{(i)}$ by solving Problem~\ref{problem:throughput_optimization_x_convex_relaxation}\;
            Derive $\bar{\vv{x}}^{(i)}$ by applying the Gaussian Randomization to $\bar{\vv{X}}^{(i)}$\vspace{1mm}
        }
        Set $\hat{\vv{x}}^{(k)}$ to $\bar{\vv{x}}^{(i)}$\;
        Derive $\hat{\vv{A}}^{(k)}$ by solving Problem~\ref{problem:throughput_optimization} given $\hat{\vv{x}}^{(k)}$
      }
\caption{Double-nested BCA algorithm for Problem~\ref{problem:throughput_optimization}}
\label{alg:throughput_optimization}
\end{algorithm}

\textbf{Inner BCA loop.} Let us start by selecting a feasible solution $\hat{\vv{A}}^{(k=0)}$ and solve the problem for $\vv{x}$. In this condition, the second addend in every element of $\vv{r}$ is non-negative and independent of $\vv{x}$ as per Eq.~\eqref{eq:objective_throughput}, so it can be neglected during this optimization stage. Therefore, we can re-write $\vv{r}$ in  Problem~\ref{problem:throughput_optimization} as $\vv{r}'$, whose elements are defined as
\begin{align}
r'_t = \sum_{j \in \mathcal{S}} \,\hat{a}^{(k)}_{tj}  \log_2  \left(1 + \frac{ g_{tj}}{g'_{tj} + N'}\right), \label{eq:objective_throughput_FP}
\end{align}
at the $k$th outer iteration, where we drop the bandwidth $W_g$, being it only a scale factor.
We tackle the above problem by decomposing it into a sequence of convex problems. First, let us define a sum-of-functions-of-ratios problem as
\begin{subequations}
\begin{align}
    & \max_{\vv{z}} && \min_t \, \sum_{j}  f_{tj}\left(\frac{\Lambda_{tj}(\vv{z})}{D_{tj}(\vv{z})} \right) & \\
    & \textup{subject to} && \vv{z} \in \mathcal{Z}, & 
\end{align}
\end{subequations}
where functions $f_{tj}(\cdot)$ are non-decreasing,  $\Lambda_{tj}(\cdot)$ are non-negative, $D_{tj}(\cdot)$ are positive and $\mathcal{Z}$ is any non-empty real set.  
By involving a sum of ratios, the above problem is a multiple-ratio Fractional Programming (FP) problem. Classical FP techniques such as Charnes-Cooper Transform or Dinkelbach's Transform cannot be readily implemented in the context of multiple-ratio FP as they would not guarantee the equality between the optimal transformed and original objective function values. 
Therefore, to decouple the numerator and the denominator of each addend of the fractional objective function, we can leverage on the \textit{Quadratic Transform}~\cite{Shen2018I}, which leads to the following equivalent formulation
\begin{subequations}
\begin{align}
    & \max_{\vv{z},\vv{y}} && \min_t \! \sum_{j}  f_{tj}(2y_{tj} \sqrt{\Lambda_{tj}(\vv{z})}\! -\! y_{tj}^2D_{tj}(\vv{z})) & \\
    & \textup{subject to} && y_{tj} \in \mathbb{R}, \quad \quad \forall t\in\mathcal{T}, j\in\mathcal{S}, \\
    & && \vv{z} \in \mathcal{Z}, & 
\end{align}
\end{subequations}
where $\vv{y}$ denotes the collection of auxiliary variables $y_{tj}$ for each fractional term $\Lambda_{tj}(\vv{z})/D_{tj}(\vv{z})$. 
We consider a particular type of FP problem, namely concave-convex, by introducing some additional hypotheses: $i$) $f_{tj}(\cdot)$ and $\Lambda_{tj}(\cdot)$ are concave functions, $ii$) $D_{tj}(\cdot)$ are convex functions, $iii$) $\mathcal{Z}$ is a nonempty convex set expressed by a finite number of inequality constraints, i.e. in standard form. In these conditions, we can solve the problem for the primary variable $\vv{z}$ and the auxiliary variable $\vv{y}$ iteratively, while dealing with a convex problem at every stage by holding one variable fixed at a time. 
Specifically, the objective function in Eq.~\eqref{eq:objective_throughput_FP} produces a concave-convex FP as $\log(\cdot)$ and $g_{tj}(\cdot)$ are concave functions and $g'_{tj}(\cdot)$ are convex functions in their domain. Thus, we proceed by developing an inner BCA loop (in contrast to the outer one addressing the optimization of variables $\vv{x}$ and $\vv{A}$), as shown in Algorithm~\ref{alg:throughput_optimization}. 
%
To this end, we apply the above-mentioned Quadratic Transform to the objective function in Eq.~\eqref{eq:objective_throughput_FP} for the $k$th step and obtain the following 
\begin{align}
r''_t\! = \!\sum_{j \in \mathcal{S}}\! \hat{a}^{(k)}_{tj}\!  \log_2\!\left(\! 2y_{tj} \sqrt{g_{tj}\! + g'_{tj}\! + \!N'}\! - \! y^2_{tj}\!\left(g'_{tj}\! + \! N'\right)\!\right)\!, \label{eq:objective_throughput_quadratic_transform}
\end{align}
which enables solving the problem for the primary variable $\vv{x}$ and the auxiliary variable $\vv{y}$ iteratively while holding one variable fixed at a time. Indeed, the problem is now convex in $\vv{y}$ for a given $\vv{x}$, thus solvable by means of convex programming, providing the optimal solution $\bar{\vv{y}}$. By substituting such value into Eq.~\eqref{eq:objective_throughput_quadratic_transform}, we need to solve the problem for $\vv{x}$, which is still non-convex in $\vv{x}$ due to nature of $\vv{b}$ in constraint~\eqref{eq:constraint_b_leq_th}. Hereafter, we take advantage of the Semidefinite Relaxation (SDR) technique to relax such constraint~\cite{luo_chang_2009}. Note that, by doing so, we derive a fully convex relaxation also of the problem in $\vv{x}$, thereby ensuring the convergence of the inner BCA loop~\cite{Grippo2000}. 
As a preliminary step, we can re-write such a constraint as follows
\begin{align}
(1 - \hat{a}'^{(k)}_t)\, \vv{\nu}_t^\intercal \vv{x} \leq  (\zeta^2_b - \hat{a}'^{(k)}_t \, u^2_t)\,\vv{x}^\intercal \vv{F}_t^\intercal\vv{x}, \label{eq:constraint_b_leq_th_least_common_multiple}
\end{align}
given that the quadratic form $\vv{x}^\intercal \vv{F}_t^\intercal\vv{x}$ is non-zero. This condition holds if there exist at least two non-zero elements of $\vv{x}$ whose corresponding sites do not share the same angular position with respect to CS $t$, which would set the squared sine of their difference to zero\footnote{The optimal solution naturally avoids such pathological conditions as the resulting PEB would explode, thus not satisfying the corresponding constraint.}. Moreover, there is no guarantee that the symmetrisation of $\vv{F}_t$ is negative semidefinite (NSD), hence Eq.~\eqref{eq:constraint_b_leq_th_least_common_multiple} is yet non-convex. 

The SDR procedure stems from the observation that $\vv{X} = \vv{x}\vv{x}^\intercal$ if and only if $X \succeq 0$ and $\textup{rank}(\vv{X}) = 1$. This observation allows linearizing the constraint as a function of $\vv{X}$, rather than $\vv{x}$. SDR drops the non-convex constraint $\textup{rank}(\vv{X}) = 1$ in order to obtain a convex relaxation of the original problem. To this aim, we point out that $\vv{x}^\intercal \vv{F}_t^\intercal\vv{x} = \Tr(\vv{F}_t^\intercal \vv{X})$ as per a well-known matrix identity, while $\vv{\nu}_t^\intercal \vv{x} = \Tr(\vv{X} \odot \vv{V}_t)$ due to the original binary nature of $\vv{x}$\footnote{Due to the original binary constraint~\eqref{eq:constraint_binary_x_a}, $x_j = x^2_j$. As $\vv{X} = \vv{x} \vv{x}^{\intercal}$, we have that $X_{jj} = x_j$.}, where $\vv{V}_t$ is a diagonal matrix such that $\diag{\vv{V}_t} = \vv{\nu}_t$. Therefore, we can write the following problem:


\begin{problem}[Inner BCA stage for $\vv{x}$]\label{problem:throughput_optimization_x_convex_relaxation}
\begin{subequations}
{\begin{align}
   & \max_{\vv{X}} && \min_{t \in \mathcal{T}} \vv{r}'' & \\
    & \textup{subject to} && \hat{\vv{A}}^{(k)} \leq \vv{1}_T \,\diag{\vv{X}}^\intercal, \label{eq:constraint_a_leq_x_SDR}\\
   &  && (1 - \hat{a}'^{(k)}_t)\, \Tr(\vv{X} \odot \vv{V}_t)\nonumber \\
   & && \hspace{5mm} \leq  (\zeta^2_b\! - \hat{a}'^{(k)}_t \, u^2_t)\!\Tr(\vv{F}_t^\intercal \vv{X}), & \forall t \in \mathcal{T}, \label{eq:constraint_b_leq_th_least_SDR} \\
   & && \Tr(\vv{X}) = G, \label{eq:constraint_cap_SDR}\\
   & && \vv{X} \succeq 0, \label{eq:constraint_x_psd}\\
   & && \diag{\vv{X}} \in [0,1]^S, \label{eq:constraint_x_psd_contin}
\end{align}}
\end{subequations}
\end{problem}
where we amend $g'_{tj}$ in Eq.~\eqref{eq:objective_throughput_quadratic_transform} as $g''_{tj} = \sum_{n \in \mathcal{S}\setminus \{j\}} g_{tn} X_{nn}$ to comply with the new optimization variable $\vv{X}$, and re-write constraints~\eqref{eq:constraint_a_leq_x} and~\eqref{eq:constraint_cap} as~\eqref{eq:constraint_a_leq_x_SDR} and~\eqref{eq:constraint_cap_SDR}, respectively.

The above problem can be solved by the interior-point method. Let us denote the solution of the $i$th step as $\{\vv{\tilde{x}}^{(i)}, \vv{\tilde{X}}^{(i)}\!\}$. Due to the SDR, the matrix $\vv{\tilde{X}}^{(i)}$ is not necessarily rank-1. If $\text{rank}(\vv{\tilde{X}}^{(i)}) = 1$, then we round $\vv{\tilde{x}}^{(i)}$ by setting its $G$ largest elements to $1$ and the remaining to $0$, and check if the result $\vv{\bar{x}}^{(i)}$ is feasible for the non-relaxed problem, i.e. for the throughput problem stage for $\vv{x}$ after the quadratic transform. If $\text{rank}(\vv{\tilde{X}}^{(i)}) > 1$ or the obtained $\vv{\bar{x}}^{(i)}$ is not feasible for the non-relaxed problem, we leverage on the Gaussian Randomization procedure to extract a feasible rank-1 component from $\vv{X}^{(i)}$. Specifically, we generate a set of $L>0$ random vectors $\vv{\xi}_l$, with $l = 1,\dots,L$, drawn from a Gaussian distribution with mean $\vv{x}^{(i)}$ and covariance matrix $\vv{X}^{(i)}$, i.e. $\vv{\xi}_l \sim \mathcal{N}(\vv{x}^{(i)},\vv{X}^{(i)})$, and project them onto the feasible region of the non-relaxed problem by setting the $G$ largest elements to $1$ and the rest to $0$ for each random sample. Then, to enforce the PEB constraint in Eq.~\eqref{eq:constraint_b_leq_th}, we remove each random sample that does not satisfy such constraint, yielding a sample size $L'\leq L$. Lastly, we choose among the obtained $L'$ feasible solutions the one $\bar{\vv{x}}$ that maximizes the objective function~\eqref{eq:objective_throughput_quadratic_transform} to be the rank-1 solution of Problem~\ref{problem:throughput_optimization_x_convex_relaxation}.

We keep executing the inner BCA loop and solve for $\vv{y}$ and $\vv{x}$, obtaining $\bar{\vv{y}}^{(i)}$ and $\bar{\vv{x}}^{(i)}$ at the $i$th inner iteration. We denote the resulting solution for $\vv{x}$ after convergence is reached as $\hat{\vv{x}}^{(k)}$, being $k$ the current outer BCA iteration.      

\textbf{Outer BCA loop.} Conversely, let us hold $\vv{x}$ fixed to the optimal value $\hat{\vv{x}}^{(k)}$ and solve Problem~\ref{problem:throughput_optimization} for $\vv{A}$ (the problem is now linear). We keep solving for $\vv{x}$ and $\vv{A}$, iterating the two outer optimization stages until convergence, and indicate the current optimal solutions as $\hat{\vv{x}}^{(k)}$ and $\hat{\vv{A}}^{(k)}$, respectively.
\begin{proposition}
\label{proposition:throughput_convergence}
The double-nested BCA algorithm for Problem~\ref{problem:throughput_optimization} converges to a stationary point.
\end{proposition}

\begin{hproof} 
According to~\cite{Grippo2000}, the convergence of BCA is guaranteed under the condition that the objective function and the feasible set are convex in each block of variables, although non-convex in general. Indeed, by means of the Quadratic Transform, the Semidefinite Relaxation (SDR) and the continuous relaxation of the binary constraints, we derive a convex relaxation of Problem~\ref{problem:throughput_optimization}. As described above, such formulation is convex in $\vv{A}$, $\vv{x}$ and $\vv{y}$ respectively, thus guaranteeing the convergence of the inner BCA loop at each outer loop iteration and then the eventual convergence of the outer BCA loop.    
\end{hproof}

\subsection{Positioning Routine}\label{s:positioning_solution}

We study Problem~\ref{problem:positioning_optimization} by partially retracing the steps performed to solve Problem~\ref{problem:throughput_optimization} and we propose an analogous routine. In particular, we make use of the Bisection Method~\cite{Boyd2004} within an analogous Block Coordinate Descent (BCD) algorithm while re-stating constraint~\eqref{eq:constraint_b_leq_th} in an equivalent form.
Indeed, straightforward algebraic manipulations show that the latter is equivalent to the following set of constraints

\begin{align}
    &\hat{a}_{tj}^{(k)}(g'_{tj} + N')(2^{\zeta_r/W_g}-1) \leq g_{tj} && \forall t \in \mathcal{T}, \forall j \in \mathcal{S} \label{eq:constraint_r_leq_th_equivalent}\\
    &\hat{a}'^{(k)}_t \left(\frac{\zeta_r}{W_e}\right) \leq c_t && \forall t \in \mathcal{T} \label{eq:constraint_r_leq_th_equivalent_lte},
\end{align}
out of which Eq.~\eqref{eq:constraint_r_leq_th_equivalent} is notably still non-convex since it contains a product of optimization variables, being $g'_{tj} = \sum_{n \in \mathcal{S}\setminus \{j\}} g_{tn} x_n$ as defined in Section~\ref{s:joint_problem}. Both loops are detailed in Algorithm~\ref{alg:positioning_optimization}.

  \begin{algorithm}[t]
    \caption{BCD algorithm for Problem~\ref{problem:positioning_optimization}}
    \label{alg:positioning_optimization}
    \DontPrintSemicolon
    Initialize $\vv{A}$ to feasible values $\hat{\vv{A}}^{(k=0)}$, $0<\eta_l<\eta_u$ and $\epsilon>0$\;
    \Repeat(\hfill\emph{Outer BCD loop (over $k$)}){convergence of objective function~\eqref{eq:objective_peb_vector} in Problem~\ref{problem:positioning_optimization}}{
        Set $\eta$ to $(\eta_l+\eta_u)/2$\;
        \Repeat(\hfill\emph{Inner Bisection loop (over $i$)}){$\eta_u - \eta_l \leq \epsilon$}{
            \uIf {Problem~\ref{problem:positioning_optimization_feasibility_problem} is feasible} {Set $\eta_u$ to $\eta$, \quad Set $\bar{\vv{x}}^{(i)}$ to its solution\;}
            \Else {Set $\eta_l$ to $\eta$\;}
        }
        Set $\hat{\vv{x}}^{(k)}$ to $\bar{\vv{x}}^{(i)}$\;
        Derive $\hat{\vv{A}}^{(k)}$ by solving Problem~\ref{problem:positioning_optimization} given $\hat{\vv{x}}^{(k)}$\;
      }
  \end{algorithm}

  \begin{algorithm}[t]
    \caption{Adaptive KPI Processing}
    \label{alg:kpi}
    \DontPrintSemicolon
    Initialize $\zeta_b \gg \max_t u_t$, $\delta$ to feasible value based on $S$ and $\epsilon > 0$\;
    Obtain $r^{(\tau = 0)}$, $\hat{\vv{x}}^{(\tau = 0)}$, $\hat{\vv{A}}^{(\tau = 0)}$ via Algorithm~\ref{alg:throughput_optimization}\;
    Derive $b^{(\tau = 0)}$ via Eq.~\eqref{eq:objective_peb} \;
    \Repeat(\hfill\emph{Thresholds tuning (over $\tau$)}){Overall \name{} convergence $\Bigm\lvert\frac{r^{(0)}-r^{*(\tau)}}{b^{(0)}-b^{*(\tau)}} - \mu \Bigm\lvert \leq \epsilon$ }{
        Set $\omega_b^{(\tau)}$ to $\frac{b^{(\tau-1)} - 1}{b^{(\tau-1)}}$,         $\zeta_b^{(\tau)} = \omega_b^{(\tau)} b^{(0)}$\;
        Obtain $r^{(\tau)}$ via Algorithm~\ref{alg:throughput_optimization}\;
        Set $\omega_r^{(\tau)}$ to $\frac{r^{(\tau-1)} - \mu}{r^{(\tau-1)}}$, $\zeta_r^{(\tau)}$ to $\omega_r^{(\tau)} r^{(0)}$ \;
        Obtain $\hat{b}^{(\tau)}$, $\hat{\vv{x}}^{(\tau)}$, $\hat{\vv{A}}^{(\tau)}$ via Algorithm~\ref{alg:positioning_optimization}\;
        Derive $\vv{x}^{*(\tau)}$ by rounding $\hat{\vv{x}}^{(\tau)}$\;  
        \For{$j \in \mathcal{S}$, $t \in \mathcal{T}$}{
        \uIf{$x^*_j = 1$ \& $\hat{a}_{tj} \leq \delta \max_t \hat{a}_{tj}$}{Set $a^*_{tj}$ to $0$}
        \ElseIf{$x^*_j = 0$}{Set $a^*_{tj}$ to $0$}}
        \For{$t \in \mathcal{T}$}{Set $a^*_{tj}$ to $1$ for $j = \underset{n|\hat{a}_{tn} \neq 0}{\argmax} \, \frac{g_{tn}}{g'_{tn}(\vv{x}^*) + N'}$ }
        Derive $r^{*(\tau)}$ via Eq.~\eqref{eq:objective_throughput} and $b^{*(\tau)}$ via Eq.~\eqref{eq:objective_peb}}
  \end{algorithm}
  
\textbf{Inner Bisection loop.} Let us describe the inner Bisection loop\footnote{Note that performing the Quadratic Transform leads to minimizing the maximum of a concave objective function, which is yet a non-convex problem.} by picking a feasible initial solution $\hat{\vv{A}}^{(k=0)}$ for the outer loop. While holding $\hat{a}'^{(k)}_t$ fixed, the second term of $\vv{b}$ is non-negative and independent of $\vv{x}$, and so it can be neglected. Indeed, we can re-write the $t$th element of $\vv{b}'$ as
\begin{equation}
   b'_t = (1 - \hat{a}'^{(k)}_t) \, \frac{\vv{\nu}_t^\intercal \vv{x}}{\vv{x}^\intercal \vv{F}_t^\intercal\vv{x}}, \label{eq:objective_peb_FP}
\end{equation}
being the denominator $\vv{x}^\intercal \vv{F}_t^\intercal\vv{x}$ non-negative. However, as discussed in Section~\ref{s:throughput}, the quadratic form $\vv{x}^\intercal \vv{F}_t^\intercal \vv{x}$ is non-convex, thus calling for the usage of SDR. The resulting function is now quasi-convex and can be tackled by means of the Bisection method. Furthermore, by keeping fixed either $\vv{x}$ or $\vv{A}$, constraints~\eqref{eq:constraint_r_leq_th_equivalent} and~\eqref{eq:constraint_r_leq_th_equivalent_lte} are convex,
leading us to state the following feasibility problem, which enables solving the inner Bisection stage for $\vv{x}$, namely 
\begin{problem}[Inner Bisection stage for $\vv{x}$]\label{problem:positioning_optimization_feasibility_problem} 
\begin{subequations}
{\begin{align}
   & \textup{Find} && \vv{X} \\
   & \textup{subject to} && (1-\hat{a}'^{(k)}_t)  \Tr(\vv{X} \odot  \vv{V}_t)  \leq \eta \Tr(\vv{F}_t^\intercal \vv{X}), \, \forall t \in \mathcal{T}, \\
   & && \eqref{eq:constraint_a_leq_x_SDR}, \eqref{eq:constraint_x_psd},\eqref{eq:constraint_cap_SDR},\eqref{eq:constraint_x_psd_contin},\eqref{eq:constraint_r_leq_th_equivalent}', \nonumber
\end{align}}
\end{subequations}
\end{problem}
where $\vv{X}$ and $\vv{V}_t$ attain to the same notation used in Section~\ref{s:throughput_solution} for the Throughput Problem, $\eta \in [\eta_l,\eta_u] \subset\mathbb{R}_+$ is the bisection threshold and constraint~\eqref{eq:constraint_r_leq_th_equivalent}$'$ is  the constraint in Eq.~\eqref{eq:constraint_r_leq_th_equivalent} compliant with the SDR formulation, by substituting $g'_{tj}$ with $g''_{tj} = \sum_{n \in \mathcal{S}\setminus \{j\}} g_{tn} X_{nn}$. Note that the Bisection Method requires exactly $\log_2((\eta_u - \eta_l)/\epsilon)$ iterations, where $\epsilon > 0$ is a given tolerance. We apply the same Gaussian Randomization procedure as aforementioned in Section~\ref{s:throughput} that returns a rank-1 solution $\bar{\vv{x}}^{(i)}$. We keep checking the feasibility of the inner Bisection stage while decreasing $\eta$ at each iteration until $\eta_u - \eta_l \leq \epsilon$. We denote the feasible solutions provided at the $i$th iteration as $\bar{\vv{y}}^{(i)}$ and $\bar{\vv{x}}^{(i)}$, respectively. 


\textbf{Outer BCD loop.} Besides, we indicate the inner loop solution for $\vv{x}$ upon convergence as $\hat{\vv{x}}^{(k)}$, where $k$ is the outer iteration. Then, we plug it into the problem for $\vv{A}$, now linear with constraints \eqref{eq:constraint_r_leq_th_equivalent} and \eqref{eq:constraint_r_leq_th_equivalent_lte}, and obtain the solution $\hat{\vv{A}}^{(k)}$. 
\begin{proposition}
The BCD algorithm for Problem~\ref{problem:positioning_optimization} converges to a stationary point.
\end{proposition}

\begin{hproof}The proof of this proposition goes analogously as the one provided for Proposition~\ref{proposition:throughput_convergence}.
\end{hproof}

\subsection{Adaptive KPI Processing and Overall Convergence} \label{s:flavour}
  
The Adaptive KPI Processing module returns thresholds $\zeta_r$ and $\zeta_b$ for each \name{} iteration $\tau$, which consists of executing both Throughput and Positioning Routines, and the quantization of relaxed variables, as clarified in Algorithm~\ref{alg:kpi}. Binary $\vv{x}^*$ is obtained by rounding $\hat{\vv{x}}$, while binary $\vv{a}^*$ is derived via the following steps in order to satisfy the constraint in Eq.~\eqref{eq:constraint_sum_a_leq_1}: $i$) set $\hat{a}_{tn}$ such that $\hat{a}_{tn} \leq \delta \,\max_t \hat{a}_{tn}$ with $n \in \{j|x_j = 1\}$ or every $\hat{a}_{tn}$ with $n \in \{j|x_j = 0\}$ to zero, where $\delta$ is finely adjusted based on $S$, $ii$) $\forall t$, set the element $\hat{a}_{tj} \neq 0$ providing the highest ratio $g_{tj}/(g'_{tj} + N')$ (given $\vv{x}^*$) to one.  

As the two routines pursue conflicting objectives in some configurations, it is unlikely that the output of one routine plugged in the other as threshold would lead to a feasible solution, thus calling for a scaling factor finely tuned to ensure the feasibility of each problem in the way to convergence. First, we run the Throughput Routine with $\zeta_b\gg \max_t{u_t}$, obtaining its solution $r^{(\tau=0)}$ along with $\vv{x}^{(\tau=0)}$ and $\vv{A}^{(\tau=0)}$ that in turn provide $b^{(\tau=0)}$ via Eq.~\eqref{eq:objective_peb}. Then, we set $\zeta_r^{(\tau)} = \omega_r^{(\tau)} r^{(0)}$, with $\omega_r^{(\tau)} = \frac{r^{(\tau-1)} - \mu}{r^{(\tau-1)}}$, and execute the Positioning Routine to get $b^{(\tau)}$, while the quantization procedure provides $r^{*(\tau)}$ and $b^{*(\tau)}$. If $\frac{r^{(0)}-r^{*(\tau)}}{b^{(0)}-b^{*(\tau)}} \simeq \mu$, convergence is reached. Conversely, we set $\zeta_b^{(\tau)} = \omega_b^{(\tau)} b^{(0)}$, with $\omega_b^{(\tau)} = \frac{b^{(\tau-1)} - 1}{b^{(\tau-1)}}$ and keep iterating until convergence. 

\section{Numerical Results}
\label{s:results}
We perform an extensive simulation campaign to test \name{} performances: we first validate our solution via Monte Carlo simulations on large instances by means of synthetic traces derived from realistic deployment settings. We then benchmark \name{} against the closest available state-of-the-art literature on a real dense-urban network topology (Bologna, Italy) provided by a major European network operator, wherein legacy (LTE) base stations are already in place and 5G-NR candidate sites are properly chosen on the map. In the following, we use terms Rate and Throughput, interchangeably. Simulation parameters are listed in Table~\ref{tab:parameters}. 
\begin{table}[t]
\caption{Simulation parameters.}
\label{tab:parameters}
\scriptsize
\centering
\resizebox{.8\linewidth}{!}{%
\begin{tabular}{c|c|c|c|c|c}
  & $\sigma_0$ & $\lambda$ [m] & P [W] & f [GHz] & W [MHz] \\  
\hline
\rowcolor[HTML]{EFEFEF}
\textbf{Legacy LTE}  & $10^{-3}$ & 10 &  30 & 1.8 &20\\
\textbf{5G-NR}  & $10^{-4}$ & 1 & 20 & 3.5&  100\\
\hline
\end{tabular}%
}
\end{table}

\begin{figure}[t]
    \centering  
    \subfigure[Highway]
    {
        \includegraphics[clip,width=0.295\linewidth]{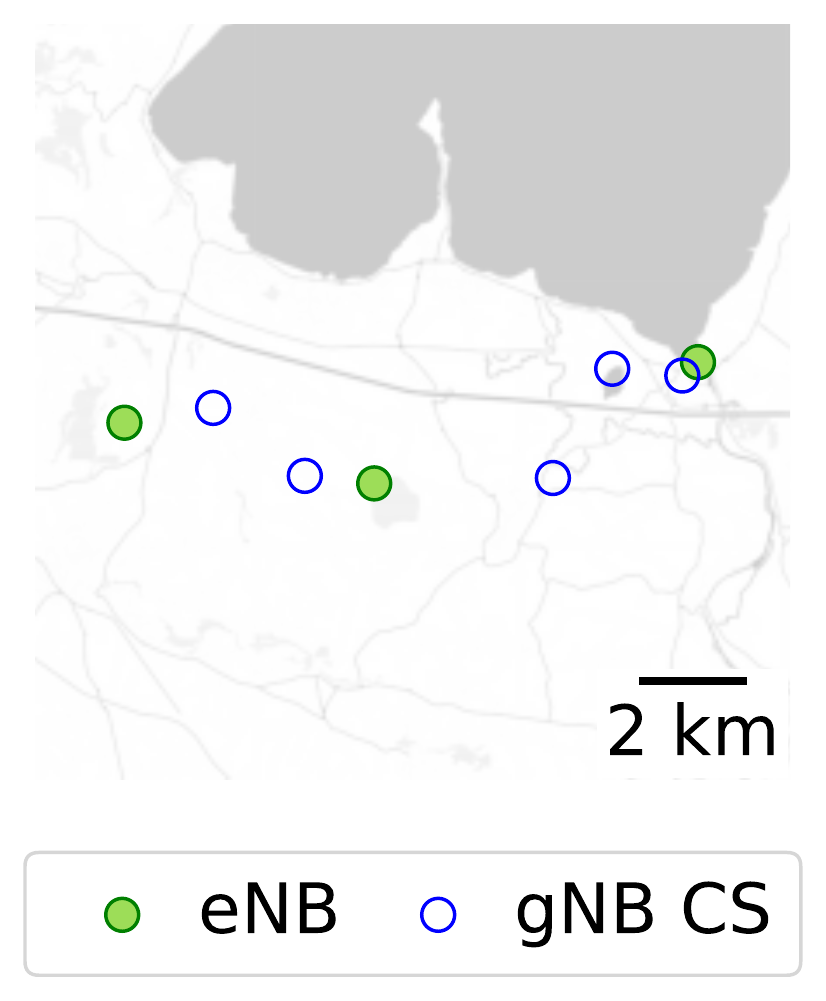}
        \label{fig:highway_scenario}
    }
    \subfigure[Suburban ]
    {
        \includegraphics[clip,width=0.295\linewidth]{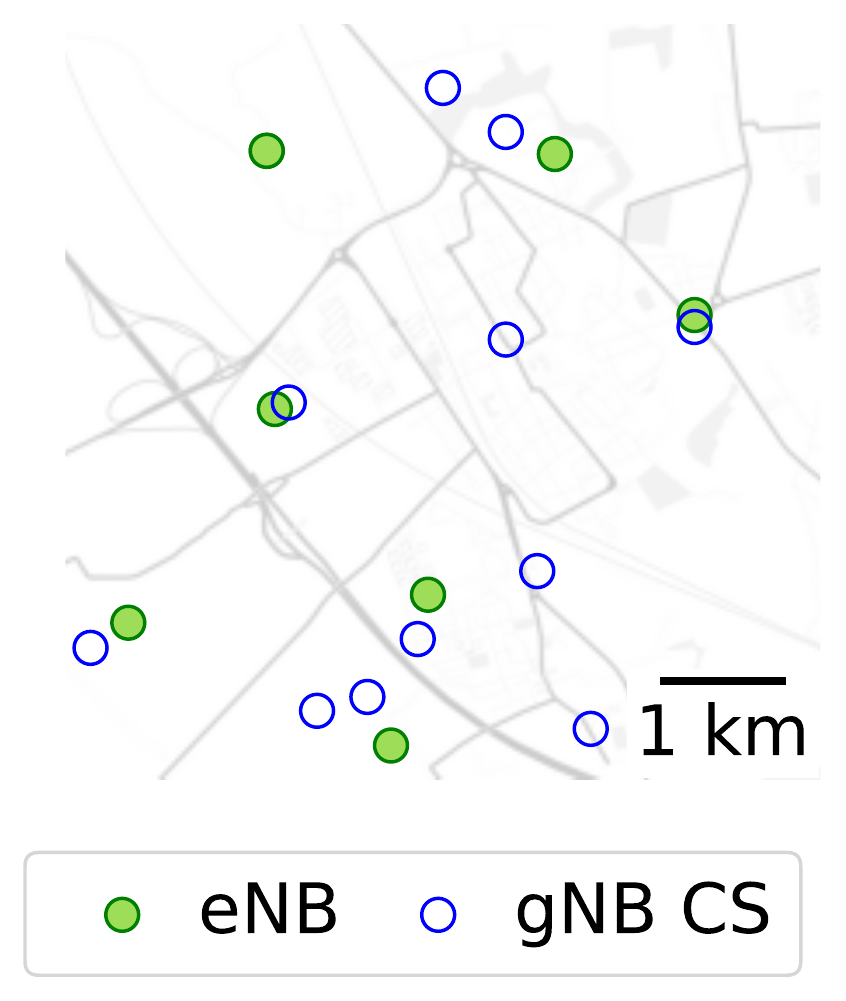}
        \label{fig:suburban_scenario}
    }
    \subfigure[Dense Urban]
    {
        \includegraphics[clip,width=0.295\linewidth]{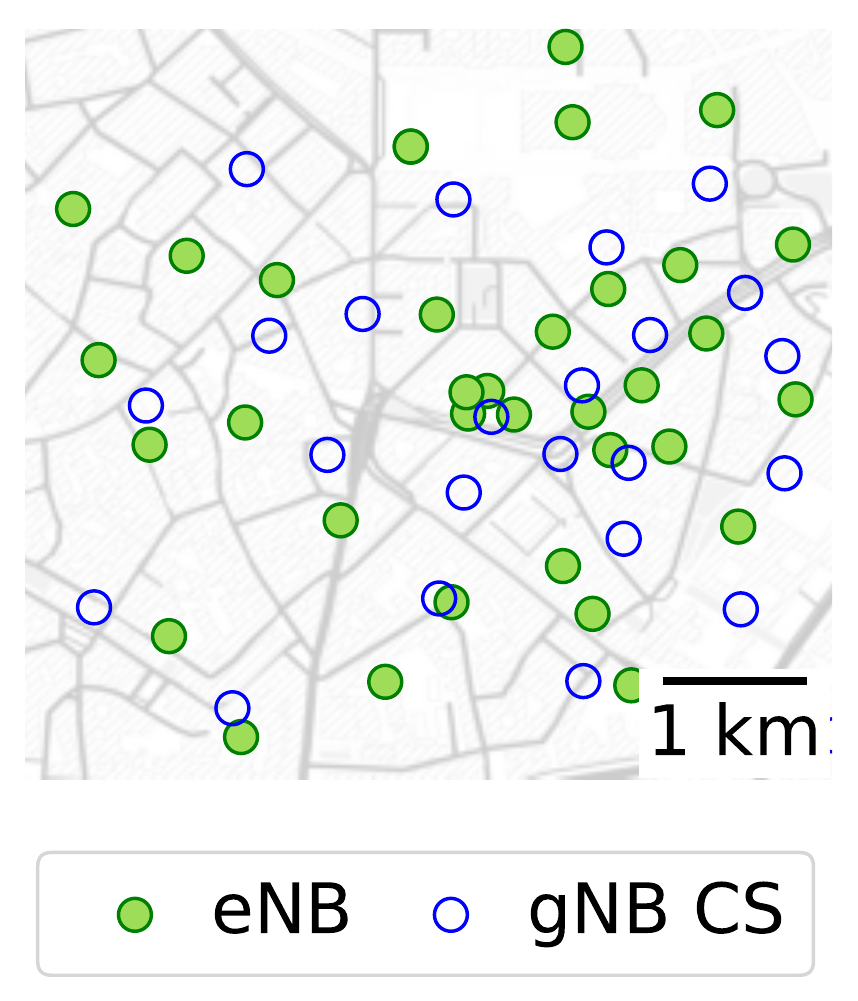}
        \label{fig:urban_scenario}
    }
    \caption{Considered scenarios based on real topologies.}
    \label{fig:scenarios}
\end{figure}

\subsection{Synthetic Topologies}
\label{s:synthetic}
We consider three different scenarios, namely Highway (H), Suburban (SU) and Dense-Urban (DU), where we place legacy base stations (eNBs) and 5G-NR (gNB) candidate sites (CSs). 
Examples are depicted in Fig.~\ref{fig:scenarios}.
For each simulation execution, we consider no deployment budget limitation, i.e. $G = S$, and randomly place eNBs and CSs while keeping fixed the base station density values obtained through realistic topologies of a major European Operator. Besides, we set propagation factor $\alpha$ and log-normal shadowing standard deviation $\sigma_s$ for each scenario as per Table~\ref{tab:parameters_scenario}.

\begin{table}[t]
\caption{Parameters by scenario.}
\label{tab:parameters_scenario}
\scriptsize
\centering
\resizebox{.65\linewidth}{!}{%
\begin{tabular}{c|c|c|c|c|c|c}
 & \multicolumn{3}{c}{Legacy LTE} & \multicolumn{3}{c}{5G-NR} \\  
 & H & SU & DU & H & SU & DU \\
\hline
\rowcolor[HTML]{EFEFEF}
\textbf{$\alpha$}  & 2.5 & 3 & 3.5 & 2.5 & 3 & 3.5\\
$\sigma_s$ [dB] & 3 & 5 & 6 & 5 & 7 & 9\\
\hline
\end{tabular}%
}
\end{table}

\begin{figure}[ht]
    \centering
    \begin{minipage}[c]{.5\linewidth}
        \centering
        \includegraphics[clip, width = \textwidth]{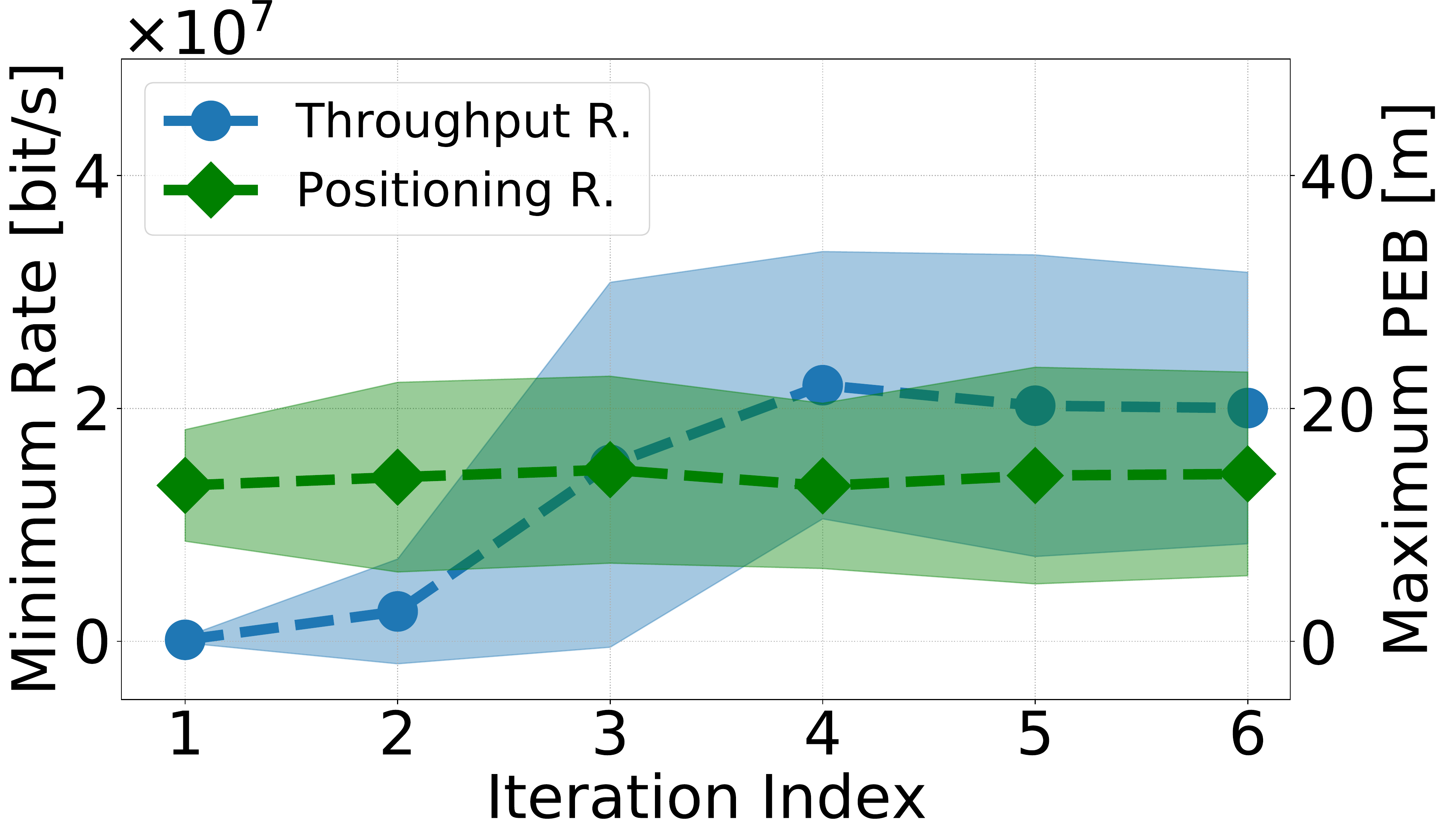}
        \captionof{figure}{Routines convergence average and standard deviation (shaded) in DU scenario.}
        \label{fig:convergence}
    \end{minipage} \hfill
    \begin{minipage}[c]{.42\linewidth}
        \hfill
        \resizebox{\linewidth}{!}{%
            \begin{tabular}[c]{c|c|c}
            \textbf{Scenario} & \textbf{\makecell{Throughput\\ Routine \\ cycles}} & \textbf{\makecell{Positioning\\ Routine \\ cycles}}\\
            \hline
            \rowcolor[HTML]{EFEFEF}
            H & 2 & 2\\
            SU & 3 & 2\\
            \rowcolor[HTML]{EFEFEF}
            DU     &  5 & 4\\
            \hline
            \end{tabular}%
            \hfill
        }\vspace{2.8mm}
        \captionof{table}{Routines convergence times.}
      \label{tab:routine_performances}
      \end{minipage}
\end{figure}

\begin{figure}[ht]
    \centering
    \subfigure[Final Throughput.]{
        \includegraphics[clip, width=0.46\linewidth ]{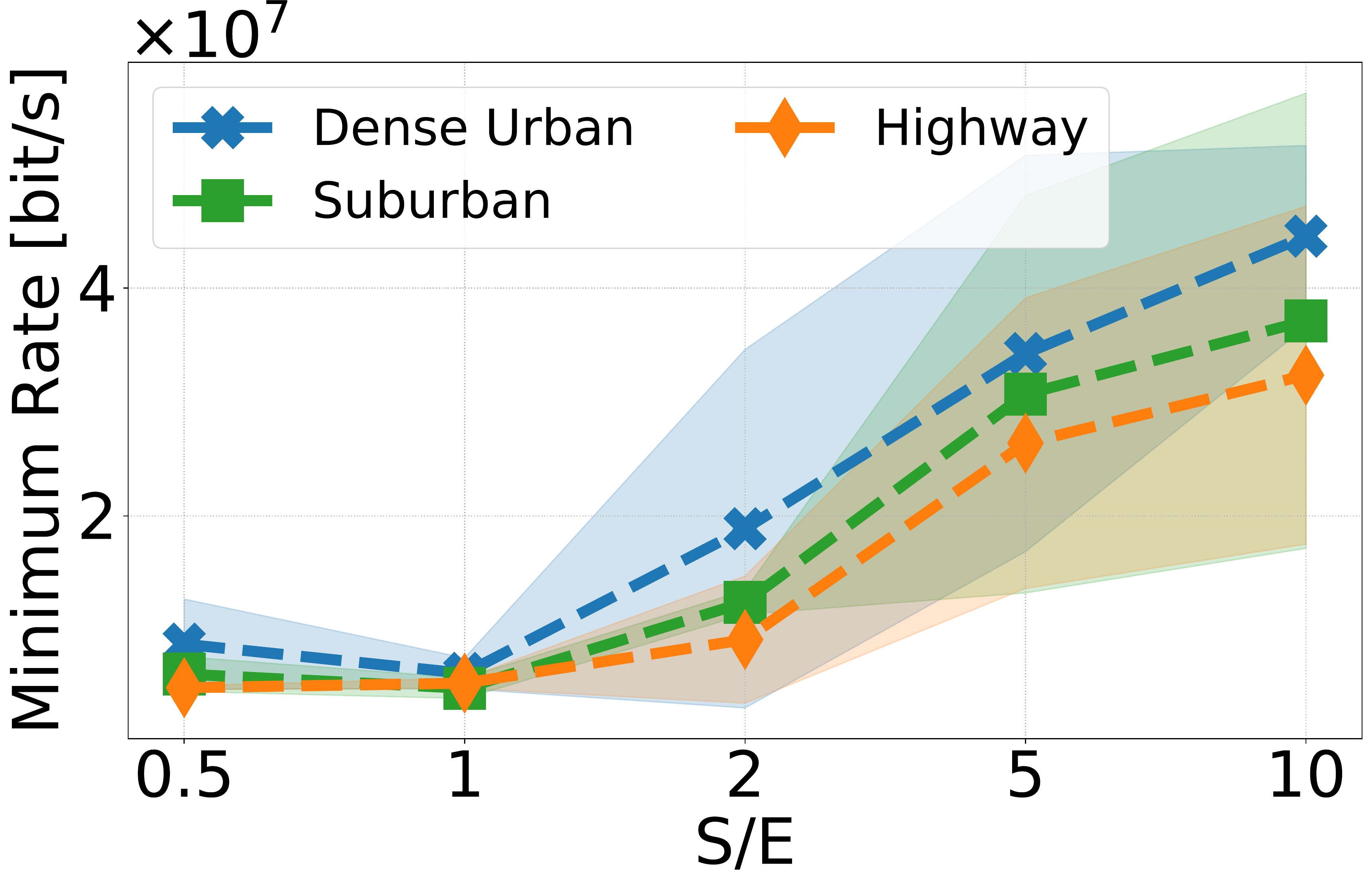}
        \label{fig:throughput_cs}
    }\hfill
    \subfigure[Final PEB.]{
        \includegraphics[clip, width=0.467\linewidth ]{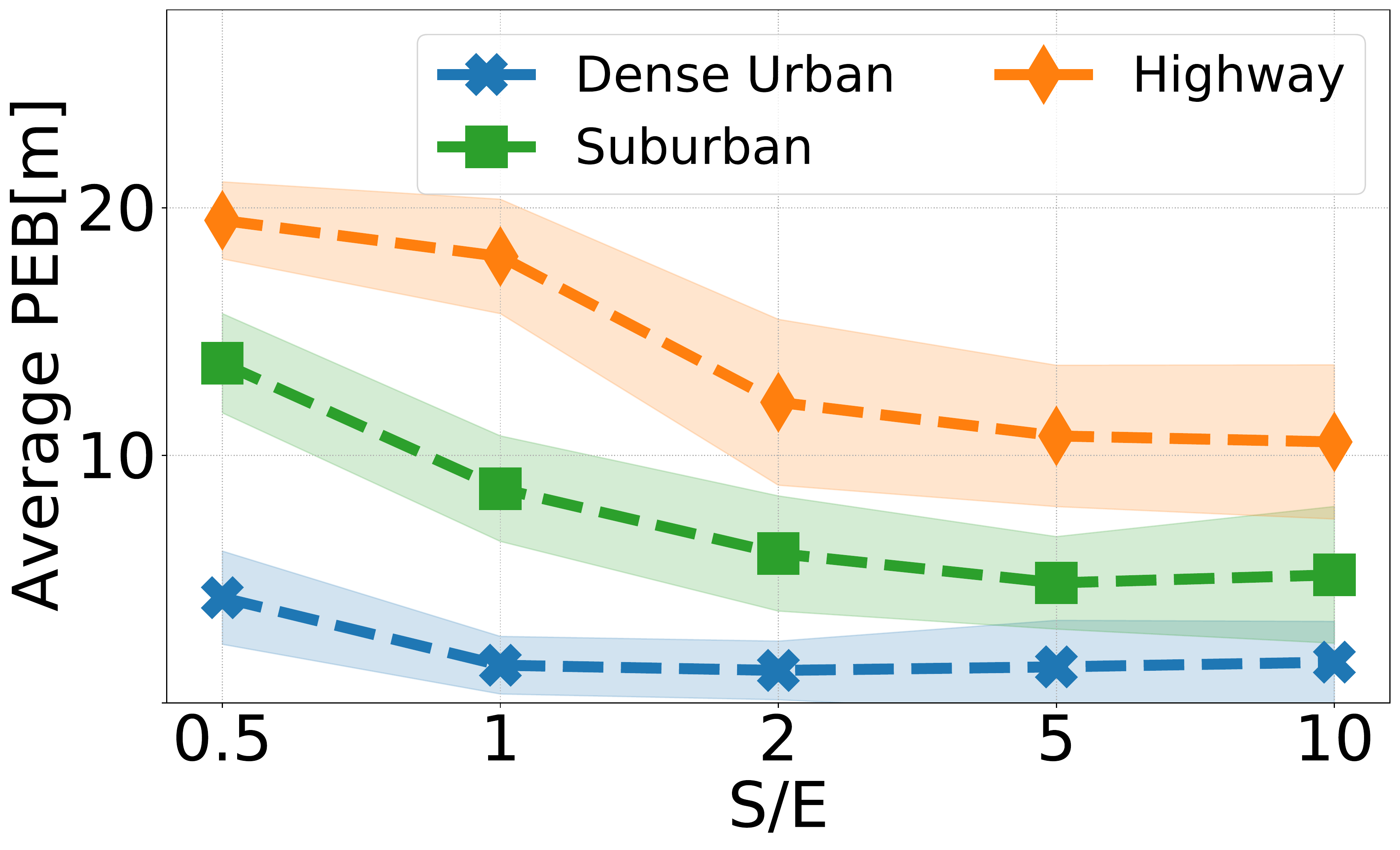}
        \label{fig:peb_cs}
    }
    \caption{\name{} performances and corresponding standard deviation (shaded) for different CS densities.}
    \label{fig:perf_cs}
\end{figure}

\noindent {\bf Convergence analysis and time complexity.} Given the relaxation procedure described in Section~\ref{s:relax}, we first validate the convergence of each individual routine. Fig.~\ref{fig:convergence} illustrates the evolution of each routine over time, showing that they converge after a few iterations. In Table~\ref{tab:routine_performances}, we list the convergence time of each routine for the various considered scenarios. As expected, while the number of eNBs and available CSs grows (see e.g., Dense-Urban scenario) the proposed algorithm takes slightly longer to converge. 
In particular, the complexity of \name{} is driven by the SDR problem in each routine, which can be solved by the interior point method with a worst-case complexity of  $\mathcal{O}(S^{5.5})$ according to \cite{Ben-Tal2001, Boyd1994}. Therefore, \name{} has greatly lower complexity than exhaustive search, which is in the order of $O(S\sum_{i,j}^{S+1} \binom{T}{i}\,\binom{S+1}{j})$. 
\change{While rapid execution might be beneficial for the computational affordability of \name{}, the main focus of our derivation is to provide a solution proved to converge regardless of the number of iterations required.}

\noindent {\bf Large-instances evaluation.} We analyze now the overall performances of \name{} increasing the ratio between available CSs and eNBs (S/E) with no deployment budget limitation, namely with $G = S$. When not explicitly mentioned, we set TPR $\mu = 10$. Moreover, we show the average PEB to reflect the maximum positioning accuracy experienced by users in the area. 
In Fig.~\ref{fig:perf_cs}, we show different performances for each selected scenario in terms of minimum throughput and average PEB. When the number of CSs increases, more degrees of freedom are introduced thus a higher granularity can be reached. However, this might require a higher CAPEX for operators that may need to install a larger number of gNBs.

\begin{figure*}[ht]
    \centering  
    \subfigure[\name{} performance in terms of minimum rate against Modified-BSE and Modified-SDR-ToA.]
    {
        \includegraphics[clip,width=.75\linewidth ]{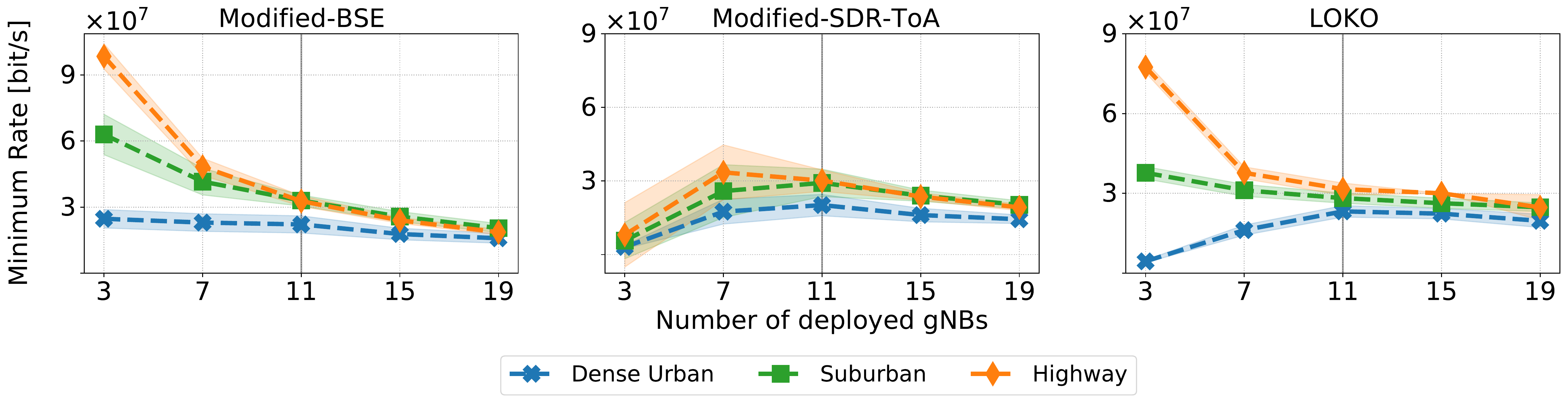}
        \label{fig:perf_num_gNBs_rate}
    }
    \subfigure[\name{} performance in terms of maximum PEB against Modified-BSE and Modified-SDR-ToA.]
    {
        \includegraphics[clip,width=0.75\linewidth]{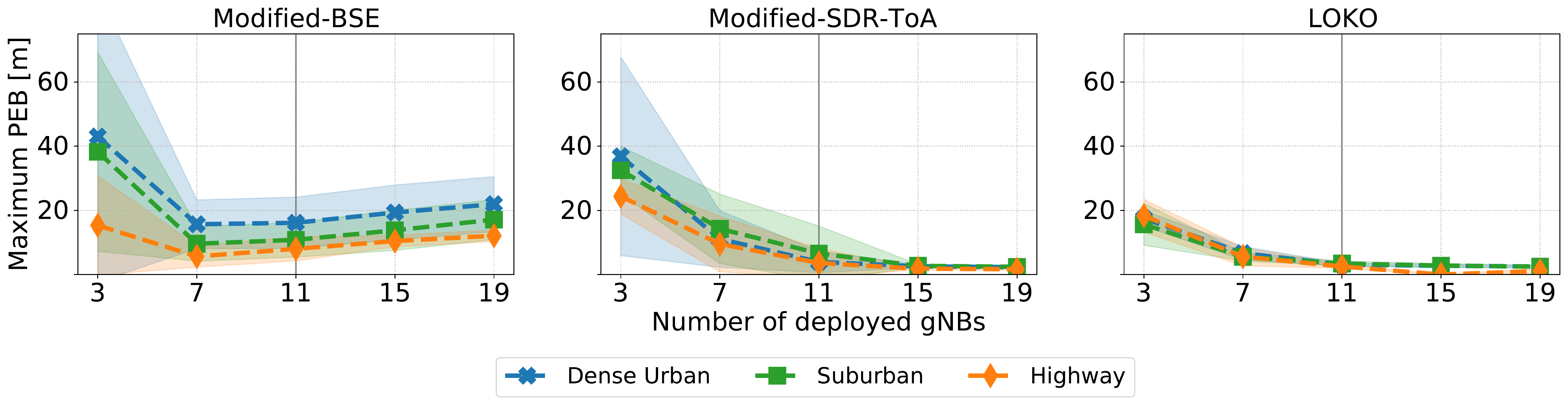}
        \label{fig:perf_num_gNBs_peb}
    }
    \caption{\name{} performance against Modified-BSE and Modified-SDR-ToA for different numbers of deployed gNBs. The vertical line indicates the number of deployed gNBs to preserve the deployment density of the reference European major network operator in each simulation scenario.}
    \label{fig:perf_num_gNBs}
\end{figure*}

\noindent\textbf{Effect of the number of deployed gNBs.} Hereafter we investigate the behavior of \name{} with respect to the number of deployed gNBs $G$ compared against the above-mentioned state-of-the-art techniques. As supported in Section~\ref{s:related}, this is the first work of its kind addressing the cellular network planning problem while optimizing user throughput, fairness and positioning KPIs. Nonetheless, to provide a significant comparison against the available literature, we hand-pick two state-of-the-art solutions, which seek the independent optimization of network throughput and localization accuracy, respectively. 

The former is dubbed as Base Station Elimination (BSE) and aims at minimizing the number of base stations covering a target area while satisfying the throughput requirements~\cite{BSE}. By default, BSE does not support multiple RATs nor a fixed number $G$ of gNBs to be deployed. Therefore, we perform some modifications of the original algorithm and design a \textit{Modified-BSE} to account for two non-interfering RATs as well as to control the number of deployed BSs. In particular, we begin with a full gNBs deployment and iteratively discard the candidate site whose corresponding gNB produces the highest total SINR (i.e., sum SINR over all test points) when eliminated in comparison with removing any other candidate site. The algorithm stops when the required number of gNBs is deployed, converging in exactly $S - G$ iterations.

The latter benchmark in~\cite{Dai2020} addresses the problem of selecting the set of anchor nodes in a dense sensor network deployment delivering the best localization performance at a given user location. Such algorithm achieves nearly optimal sensor selection by minimizing the trace of the CRLB, whose direct relationship with the PEB is described in Section~\ref{s:ToA}, under the condition of fixed number of selected sensors. As the original formulation is not bespoke for the purpose of cellular network deployment, we introduce a slight modification of its objective, namely we optimize the PEB fairness among all UEs locations by minimizing the maximum of the above-mentioned CRLB trace over all test points. It is worth pointing out that this solution complies with the widely used simplifying assumption that ToA measurement errors are independent of the actual distance, which translates into a fixed-variance Gaussian noise. \change{Therefore, it optimizes for a PEB formulation that accounts only for the angular positions of the CSs with respect to the test points, which is not guaranteed to provide good performance when evaluating the resulting PEB as per Eq.~\eqref{eq:objective_peb}.} We refer to this amended version of the algorithm as \textit{Modified-SDR-ToA} approach.\footnote{Modified-SDR-ToA has a complexity of $\mathcal{O}(S^{4.5})$ while BSE has a complexity of $S^2\mathcal{O}((S-1)T)$.} Furthermore, the exhaustive search method is introduced in the performance benchmark as a greedy approach if computationally affordable for the specific problem instance.

Fig.~\ref{fig:perf_num_gNBs} illustrates the throughput and localization performances of \name{} compared against the above-mentioned benchmarks for different numbers of deployed gNBs. We consider a range of gNBs such that the middle point $G = 11$, indicated by a vertical line, returns the same BS density as the one provided by our reference major European operator. As for the minimum rate depicted in Fig.~\ref{fig:perf_num_gNBs_rate}, Modified-BSE has a monotonic decreasing behavior while Modified-SDR-ToA has an approximately concave trend. Indeed, the latter technique attempts at optimally selecting candidate sites in the service area so as to improve upon the minimum PEB provided by the LTE network. However, the 5G network proves incapable to deliver better localization performance than LTE in case only a handful of gNBs are deployed, thereby forcing several test points to be served by the LTE network under the assumption of hybrid BS-UE association, as illustrated in Section~\ref{s:joint_problem}. Increasing the number of gNBs, the 5G deployment starts outmatching the LTE PEB and moves more and more test points to the 5G network, resulting in a bump in throughput performance in the middle gNBs range. 
Notably, \name{} has a cross-breed behavior, which traces Modified-BSE in the highway and suburban scenarios, and Modified-SDR-ToA in the urban scenario, in the low number of deployed gNBs regime. In particular, \name{} is able to deliver higher throughput with a few available gNBs whenever the PEB constraint is comparatively easier to satisfy, e.g., in the H and SU scenarios, thanks to the inherently higher spatial sparsity of BSs with respect to the test points in bigger deployment areas.

\begin{figure}[t!]
    \centering  
    \subfigure[\name{} performance against BSE for the same number of deployed gNBs as obtained at BSE convergence]
    {
        \includegraphics[clip,width=.81\linewidth ]{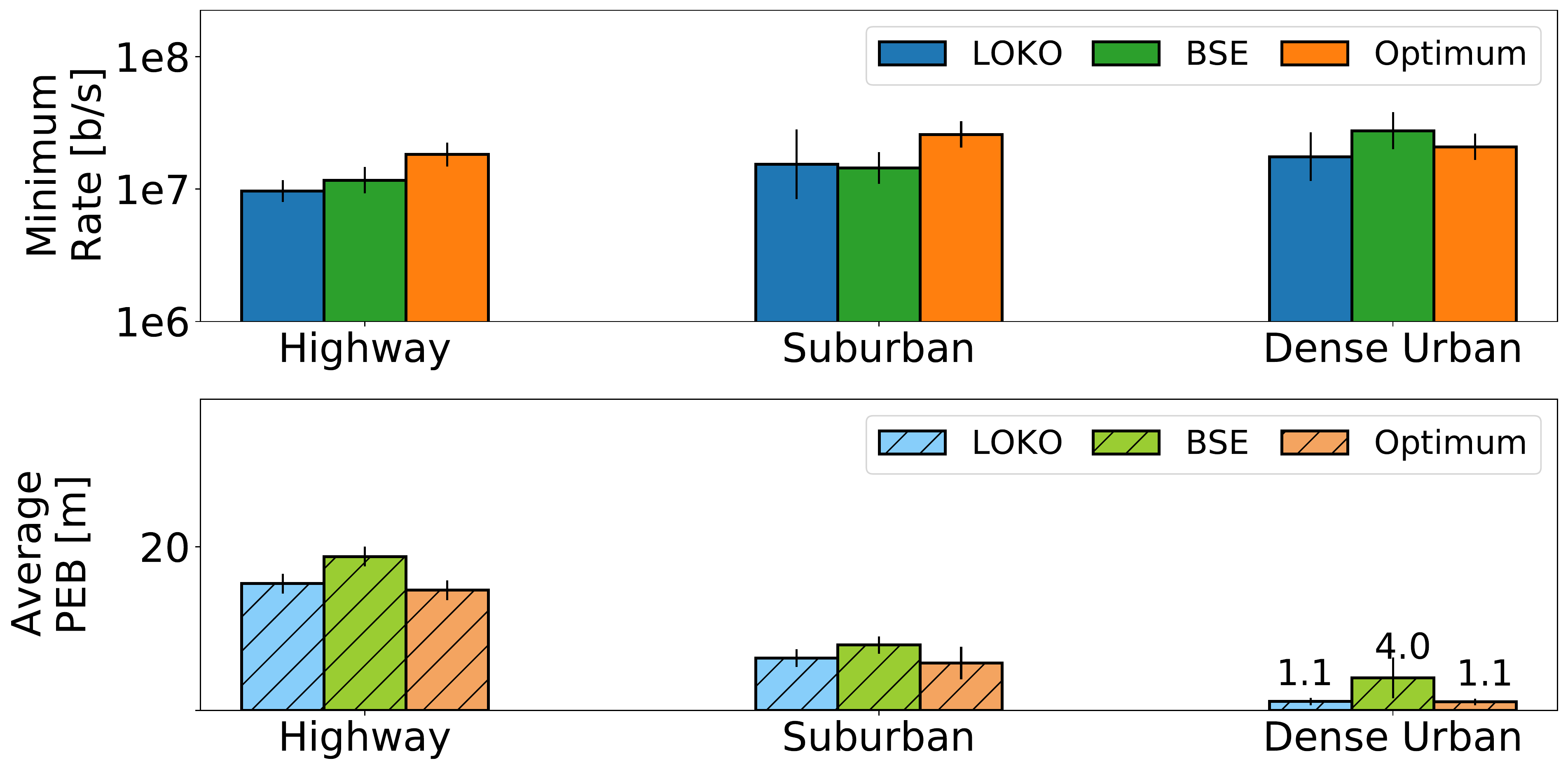}
        \label{fig:barplot_throughput}
    }
    \subfigure[Number of deployed gNBs by \name{} against BSE when delivering the same average PEB performance as BSE upon its convergence.]
    {
        \includegraphics[clip,width=0.81\linewidth]{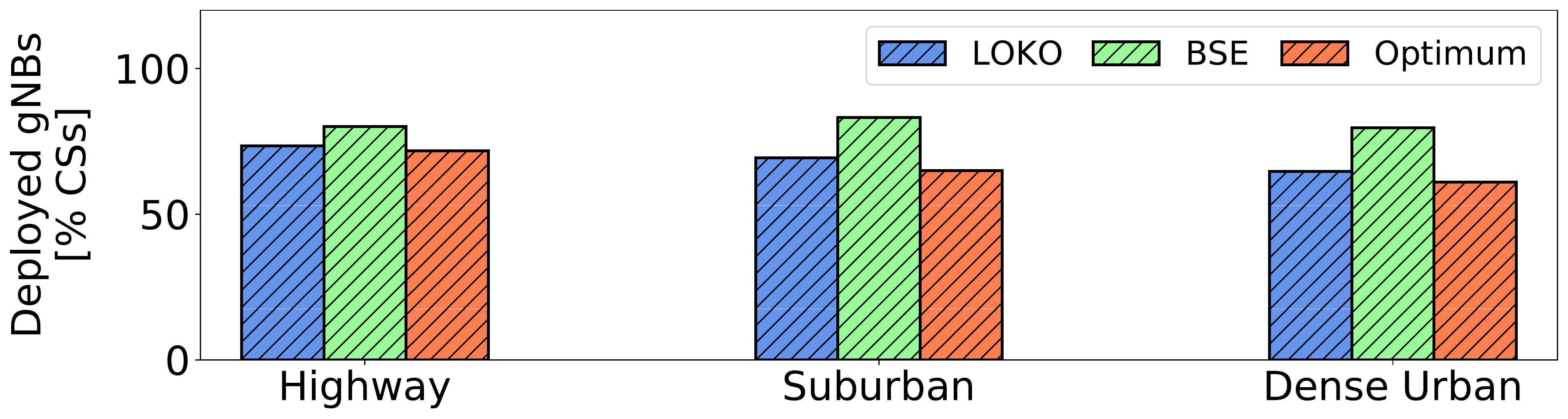}
        \label{fig:barplot_peb}
    }
    \caption{\name{} CAPEX efficiency against the BSE algorithm (c.f.~\cite{BSE}) and the optimum derived via exhaustive search.}
    \label{fig:barplot}
\end{figure}

On the other hand, the analysis in Fig.~\ref{fig:perf_num_gNBs_peb} of the localization performance shows the great advantage of \name{} against its competitor solutions. As expected, Modified-BSE provides the worst results but even Modified-SDR-ToA is outmatched by as much as $60\%$ (reduction in PEB w.r.t. the value obtained by Modified-SDR-ToA in low gNBs regime). The reason behind it is that Modified-SDR-ToA does not consider the measurement error dependency on the actual distance from the anchor BS, therefore calculating sub-optimal UE-BS associations. We would like to highlight that \name{} delivers similar performance for all scenarios in terms of maximum PEB, which is likely provided by the LTE network. Nonetheless, these results prove that \name{} is able to deploy gNBs as well as establish UE-BS associations without letting the majority of UEs connect to the less-performing LTE network by providing an effective 5G deployment even with a few available gNBs to be deployed.

\noindent\textbf{CAPEX utilization efficiency.} Inspired by the original BSE~\cite{BSE}, which does not assume any constraint on the number of gNBs, we test the efficiency of \name{} in terms of CAPEX utilization. Specifically, we consider the same scenarios (H, SU, DU) and run BSE until convergence. Then, we run \name{} and set $G$ equal to the number of deployed gNBs upon BSE convergence. Note that BSE convergence happens at different number of iterations in each scenario. In Fig.~\ref{fig:barplot_throughput}, we depict the corresponding performance in terms of minimum rate and average PEB. Interestingly, \name{} significantly outperforms BSE (about $-75\%$) in terms of PEB while showing similar throughput results (about $-10\%$ on average) when the same number of gNBs is utilized, achieving near-optimal results at an affordable time complexity. In addition, we report in Fig.~\ref{fig:barplot_peb} the fraction of deployed gNBs with respect to the overall CSs such that \name{} and BSE approximately deliver the same localization performance, proving that \name{} has a lighter footprint on the available resources in almost all tested scenarios. There results unveil that optimally planning the deployment at the available sites is crucial to pursue joint throughput and localization performance maximization. Moreover, they greatly motivate the adoption of \name{} for dense-urban scenarios where geographical constraints require a complex but practical solution to reach the expected performances.

\begin{figure}[t]
    \centering
    \includegraphics[width=\linewidth ]{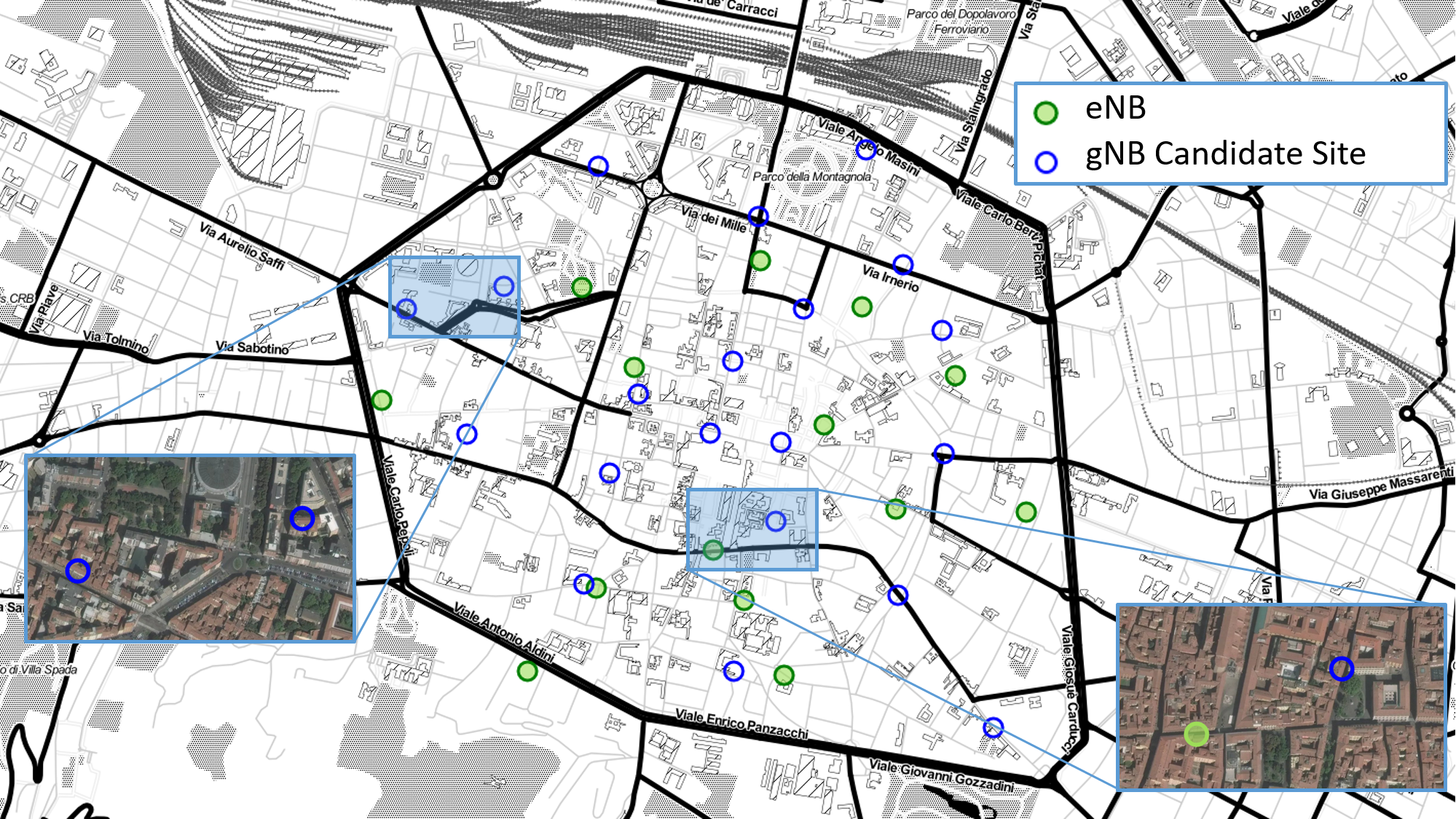}
    \caption{Legacy (LTE) deployment and selected 5G-NR CSs.}
    \label{fig:map}
\end{figure}

\begin{figure}[t]
    \centering
    \subfigure[LTE Throughput]
    {
       \centering
        \includegraphics[clip,width=0.46\linewidth]{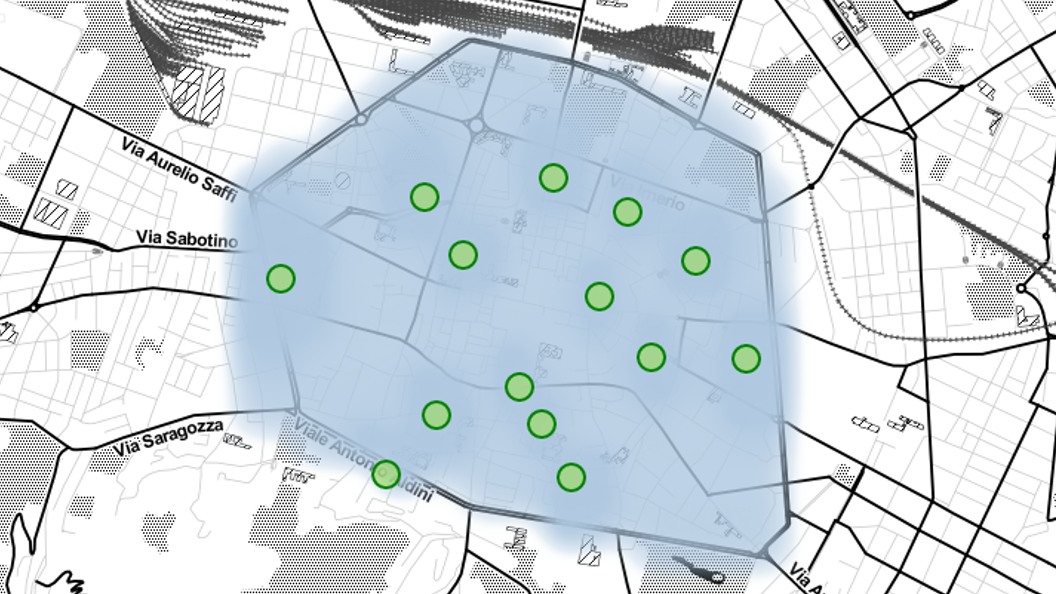}
        \label{fig:enb_throughput_heatmap}
    }
    \subfigure[\name{} Throughput]
    {
       \centering
        \includegraphics[clip,width=0.46\linewidth]{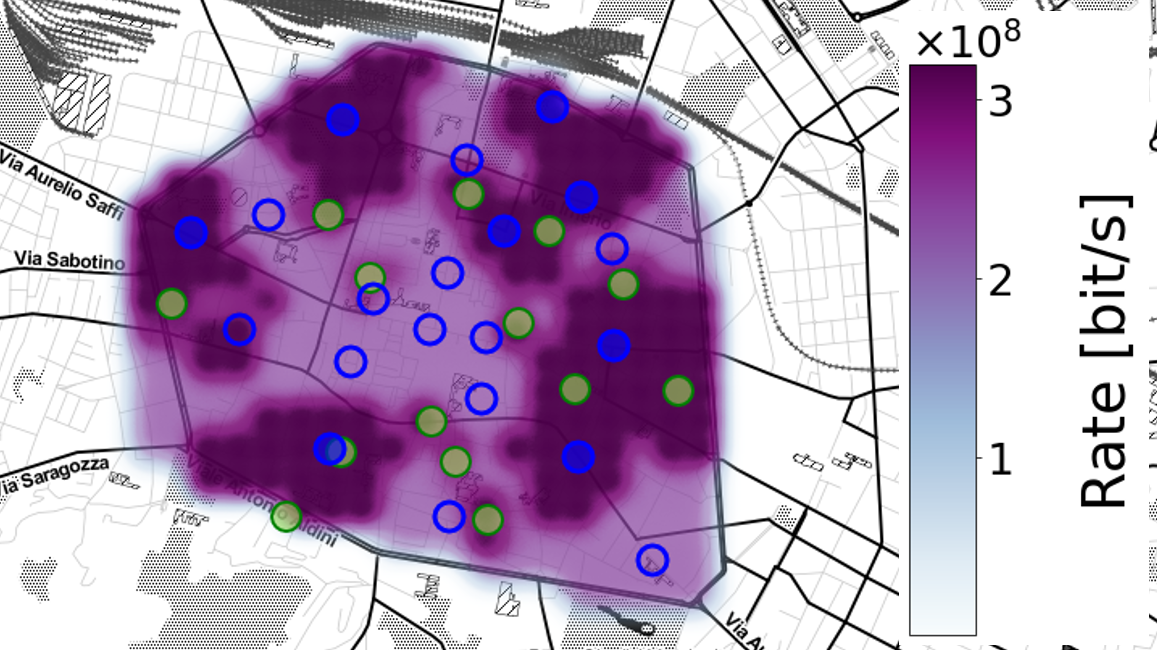}
        \label{fig:loko_throughput_heatmap}
    }
    \subfigure[LTE PEB, Average $=45$ m]
    {
       \centering
        \includegraphics[clip,width=0.46\linewidth]{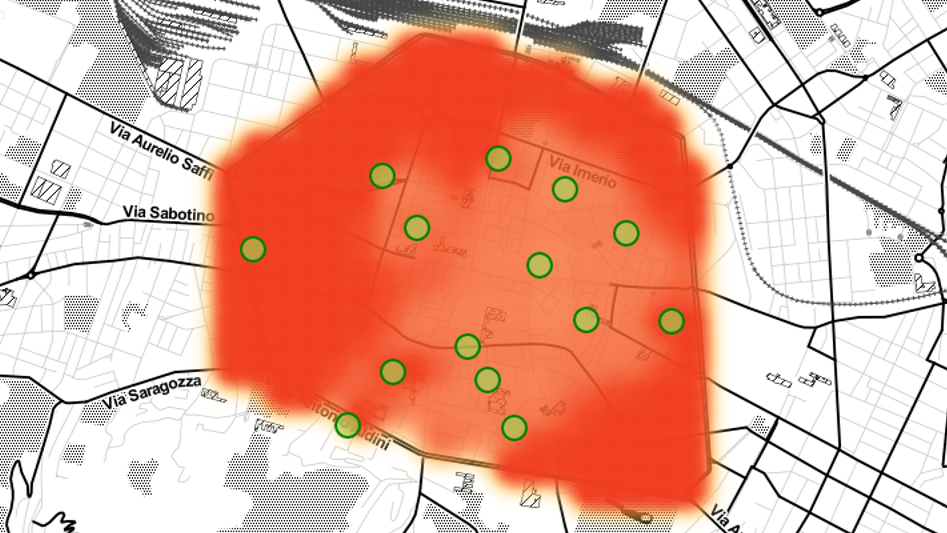}
        \label{fig:enb_peb_heatmap}
    }
    \subfigure[\name{} PEB, Average $=1.7$ m]
    {
       \centering
        \includegraphics[clip,width=0.46\linewidth]{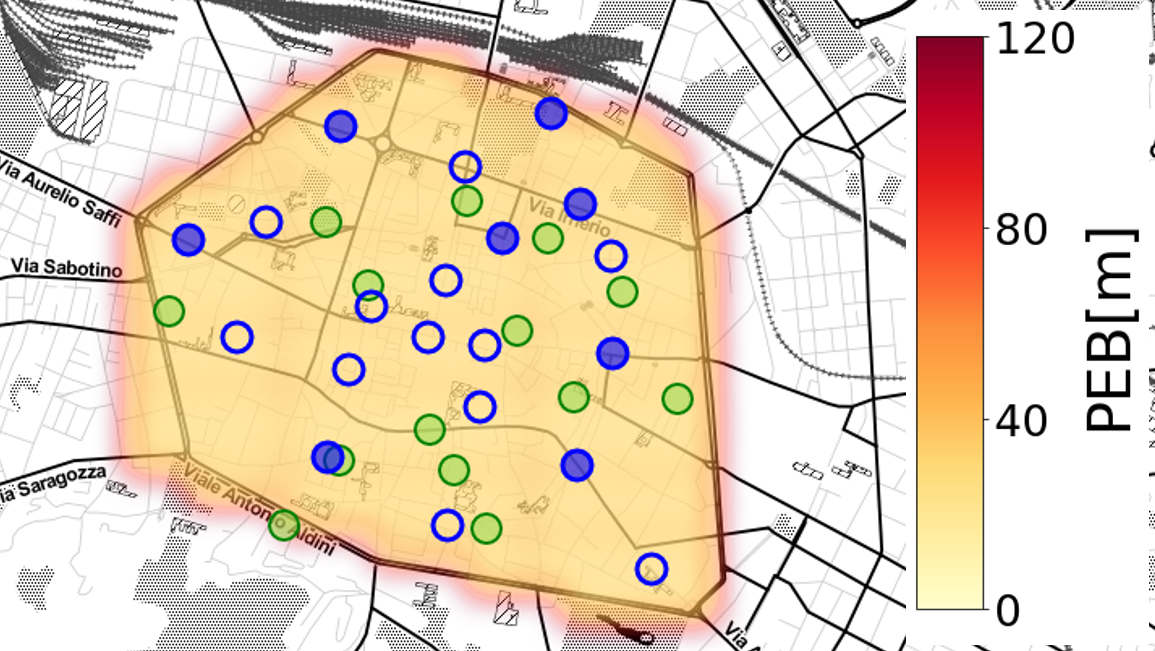}
        \label{fig:loko_peb_heatmap}
    }
    \caption{Final results on Bologna Ringway.}
      \label{fig:bologna_heatmap}
\end{figure}

\subsection{Realistic Deployments}
\label{s:real_depl}
Dense-urban scenarios fully exploit the ability of our solution to correctly balance off between throughput and positioning performances. Thus, we use a live-network dataset from a major European operator considering eNBs deployed within the Bologna Ringway~\cite{Bologna_ringway}. We add a new tier that includes 5G-NR CSs, properly selected based on the operator guidelines from the Bologna map, as shown in Fig.~\ref{fig:map}.

We apply our solution to choose the minimum number of locations among $20$ available CSs where to install gNBs. In Figs.~\ref{fig:enb_throughput_heatmap} and \ref{fig:enb_peb_heatmap}, we show the performances of the legacy LTE deployment in terms of Throughput and PEB. \name{} deployes $8$ gNBs and improves such performances by about $100\%$ and $90\%$ respectively, as shown in Figs.~\ref{fig:loko_throughput_heatmap} and \ref{fig:loko_peb_heatmap}, partly due to 5G-NR boosted efficiency.

Finally, our solution enables operators to decide whether throughput performance has a higher priority w.r.t. PEB in the optimization process by means of TPR $\mu$. Different values of TPRs and obtained results are summarized in Table~\ref{tab:tpr}. Note that when $\mu=0$, \name{} optimizes only the minimum throughput whereas when $\mu=20$ it mostly optimizes the PEB. We compare \name{} against Modified-BSE and Modified-SDR-ToA. Results show that \name{} already substantially outperforms Modified-BSE and Modified-SDR-ToA in terms of PEB even for $\mu = 0$, and further improves PEB as $\mu$ grows. \change{As expected, throughput performance decreases as $\mu$ increases. For $\mu = 0$, \name{} provides a throughput performance comparable to BSE and M-SDR-ToA. As both Throughput and Positioning Routines need to be executed at least once before reaching convergence, even for $\mu = 0$, \name{} goes through the optimization of the PEB-constrained throughput and the throughput-constrained PEB, respectively. This has an impact on the final output, which results in a reduction of the minimum throughput of $0.6 \%$ w.r.t. M-SDR-ToA, which is well within the effect of numerical artifacts introduced by SDRs and related Gaussian randomization procedures. The deployment provided by M-SDR-ToA, by selecting CSs so as they are as much uniformly spread as possible around each test point, generates \textit{as byproduct} good throughput performance in the considered dense urban environment, which is likely interference-limited. This is in full accordance with the results obtained on synthetic network topologies, as shown in Section~\ref{s:synthetic}. 
}

\begin{table}[t]
\caption{Final results on Bologna Ringway against TPR.}
\label{tab:tpr}
\scriptsize
\centering
\resizebox{\linewidth}{!}{%
\begin{tabular}{c|c|c|c|c|c|c}
\textbf{TPR} &  \textbf{M-BSE} & \textbf{M-SDR-ToA} & \textbf{0} & \textbf{5} & \textbf{10} & \textbf{20}\\  
\hline
\rowcolor[HTML]{EFEFEF}
\textit{Min. Throughput} [Mbit/s]  & $53.6$ & $50.4$ & $50.1$ & $45.5$ & $33.5$ &  $5.6$\\
\textit{Maximum PEB} [m] & $18.6$ & $16.7$ & $14.1$ & $12.9$ & $12.4$ & $11.9$ \\
\rowcolor[HTML]{EFEFEF}\textit{Average PEB} [m]  & $13.1$ & $11.5$ & $7.1$ & $5.8$ &$1.7$ &$1.2$ \\
\hline
\end{tabular}%
}
\end{table}

\section{Related Work}
\label{s:related}

\textbf{Radio Frequency (RF) Planning.} The problem of selecting where to install base stations is fundamental since the first cellular network deployments~\cite{Hurley2002,Amaldi2003}. The typical KPI for such works is coverage probability~\cite{Huang2013,Andrews2011,Koutitas2012}, which is shown to be almost invariant with respect to base station location geometry in case of massive deployments~\cite{Brown2000}. However, the advent of 5G and its advanced radio technology has led to a paradigm change~\cite{pollution5G2020,Penttinen2019,Chiaraviglio2018,albanese2022} and brought up a series of new challenges, such as the usage of millimeter wave radio transceivers~\cite{Andrews2017} and ultra-dense non-uniform deployment~\cite{Rezaabad2018,Lopez-Perez2015} in which case a number of heuristic algorithms have been developed to optimize both coverage and throughput~\cite{Wang2014,Dong2015}. To the best of our knowledge, this is the first work that considers localization accuracy in the network planning problem. 

\noindent \textbf{Localization Error bound.} Providing a definition of localization error is crucial when dealing with positioning systems. The most used indicator is the CRLB, which has been derived neglecting measurement biases due to NLoS propagation conditions in early works like~\cite{Ho2007,Ho2004}. Conversely, more recent works address the measurement biases as additional estimation parameters~\cite{Qi2006,Jourdan2008}, which in turn improve the localization accuracy only if some prior information about them, e.g., their probability density function, is available~\cite{Han2016,Win2018}. The adoption of mmWave radio technology has opened up new possibilities owing to their high channel angular sparsity, which allows estimating position-related parameters for each recevied NLoS path. For instance, \cite{Fascista2019} solves the UE localization problem for a multiple-input single-ouput (MISO) system, addressing the cases where massive arrays are deployed only at the base station side due to initial technical limitations integrating them into compact UEs. The bounds for 3D localization via mmWave are derived in~\cite{Abu-Shaban2018}, showing their theoretical capability of localizing a UE with sub-meter position error and sub-degree orientation error, though with differences in availability accuracy between uplink and downlink. Besides, mmWave MIMO can provide a progressive map of the environment by solving the so-called simultaneous localization and mapping (SLAM) problem, e.g. leveraging on diffuse multipath~\cite{Wen2021}. In our work, we have leveraged on a CRLB formulation that takes into account the dependency of the estimation error on the actual distance between the UE and the BS, thereby being general and able to capture the performance of all ToA-based techniques in the network planning phase.

\noindent \textbf{Sensor Selection Problem.} User positioning has been widely investigated in the literature and is still an active research track being of practical importance for many applications in wireless communications, ranging from navigation and radar systems to cellular and wireless sensor networks (WSNs)~\cite{Ko2015,Nordio2015,Cao2016}. Specifically, it often translates to a sensor selection problem with different objectives, such as energy efficiency, connectivity robustness, positioning accuracy and more. Various works tackle these problems with heuristic approaches under the hypothesis of an uncorrelated or weakly correlated measurement model~\cite{Chepuri2015,Shen2014}, which can be efficiently handled by means of convex optimization. However, this assumption does not hold in real scenarios where sensor measurements are likely correlated~\cite{Jindal2006}. Therefore, the resulting estimation accuracy is a non-linear function of the sensor selection variables~\cite{Liu2016}, thus leading to a non-convex optimization formulation. Although greedy algorithms have been proposed~\cite{Shamaiah2010}, convex relaxation methods involving SDR provide greater performances, approaching exhaustive search at least on limited-size instances~\cite{Chepuri2015,Zhao2019,Dai2020}. Besides the apparent difference in technological domain between WSNs and cellular networks, sensor selection problems do not generally aim at optimizing network throughput as the achievable communication rate is not a compelling parameter for sensor nodes. As a result, these works end up with problem formulations that are fundamentally different from ours.

To the best of our knowledge, there is not any prior literature addressing the joint optimization of network throughput and positioning performances in the network deployment process.  
\section{Conclusions}
\label{s:conclusions}
In this paper we proposed, analysed and evaluated \name, a novel method for 5G roll-out planning that, in contrast to state-of-the-art solutions, considers \emph{mobile users localization accuracy requirements}. \name{} is designed to address the needs of service providers when considering \emph{where to place 5G base stations} to maximize  throughput performances, taking into account \emph{mobile users localization targets}.
In order to accomplish this, we defined a new configuration parameter, \emph{Throughput-Positioning Ratio (TPR)}, which enables operators to configure customized policies towards achieving a given positioning accuracy. \name{} iteratively relies on Block Coordinate, Semidefinite Relaxation (SDR) and Bisection techniques to find the right location for 5G base stations to meet predefined localization requirements within a given \emph{available candidate sites set} and/or geographical area. Our results, considering \emph{real base station deployments}, demonstrated the effectiveness of the solution along with the trade-offs between throughput and localization to be expected. Overall, in the large-scale scenarios considered, \name{} achieved up to one order of magnitude  PEB improvements (tenths of meters to meters) depending on the \emph{TPR} configuration. 

\section{Acknowledgement}
This work was supported by EU H2020 RISE-6G (grant agreement 101017011).


\bibliographystyle{IEEEtran}
\bibliography{bibliography}

\end{document}